\documentclass[aps,physrev,reprint,groupedaddress]{revtex4-2}

\usepackage{graphicx,url}
\usepackage{bm}
\usepackage{siunitx,upgreek}
\usepackage[version=4]{mhchem}
\usepackage{textgreek}
\usepackage[caption=false]{subfig}
\begin{document}
\title{Critical nonreciprocity in gyrotropic coupled-waveguide system for TE-mode optical isolator and circulator}
\author{Kimhong Chao}
\author{Vy Yam}
\author{Laurent Vivien}
\author{Béatrice Dagens}
\affiliation{Université Paris-Saclay, CNRS, Centre de Nanosciences et de Nanotechnologies, 91120 Palaiseau, France}

\date{\today}

\begin{abstract}
Photonic integrated circuits (PICs) increasingly require more advanced integrated functionalities and devices to meet the key application challenges. This evolution with the serial integration of optical functions in the same photonic circuit often results in internal reflections and optical feedback, which can specifically destabilize lasers. One solution to overcome this issue is to integrate an optical isolator at the output of the lasers. Among various proposed designs, magneto-optical-based isolators have significantly improved over the past decades in terms of compactness, insertion loss, isolation ratio, and spectral isolation bandwidth. Despite these improvements, the TE optical isolator still lacks in performance. This paper introduces a new operational principle for a TE optical isolator based on modal beating in a transverse magneto-optical Kerr effect (TMOKE) coupled-waveguide system. This approach combines evanescent-coupled silicon waveguides and the magneto-optical effect, resulting in a nonreciprocal propagation that can be optimized for optical isolation. This new concept shows promise in achieving a high-performing TE optical isolator, as it does not depend on resonance and is free from constraints associated with interferometers. Based on data from magneto-optical garnet materials, the simulated device has a length of \qty{500}{\um} and its \qty{20}{dB}-isolation bandwidth is as high as \qty{35}{nm}. With broadband and high isolation, this simulated device opens possibilities for miniaturizing complex photonic circuits used in optical communication, data communication, and optical sensing.
\end{abstract}


\maketitle

\section{Introduction}
\label{sec_intro}

The optical isolator is a crucial component in photonics, as it introduces nonreciprocity to the optical system. This feature helps prevent unwanted disturbances caused by optical feedback in active components, such as laser sources \cite{petermanExternalOpticalFeedback1995}. However, integrating a laser source and an optical isolator on a chip remains a significant challenge despite advancements made over the past few decades. This issue presents a scalability problem for PICs \cite{shekharRoadmappingNextGeneration2024}, as the current solution typically involves using an off-chip approach where the laser is connected to the PIC through fiber or lens attachment along with a bulk optical isolator. This setup increases packaging complexity and cost. To further improve the scalability of the PICs, on-chip solutions to laser integration with reflection control have been a widely discussed research topic \cite{harfoucheKickingHabitSemiconductor2020, gomezHighCoherenceCollapse2020, duan13$mu$Reflection2019,huangFeedbackTolerantQuantum2024}.

The current off-chip isolator is based on magneto-optical (MO) material, which uses Faraday rotation and includes \ang{45}-offset polarizers at both ends to achieve isolation. While this method is effective in free-space optics, miniaturizing this device on an integrated platform is difficult due to the modal birefringence of the waveguides and the incompatibility of the usual MO garnet and the integrated platform. Several alternative solutions have been proposed to address this challenge, including MO-free options such as spatiotemporal modulators \cite{tianMagneticFreeSiliconNitride2021} and active reflection cancellation circuits \cite{shomanStableReducedLinewidthLaser2021a}. Although these devices exhibit the nonreciprocal effect, which favors optical isolation, their performance is not as good as that of off-chip isolators, and they tend to add more complexity to the design.

However, despite material challenges, there have been significant advancements in MO-based integrated optical isolators. These developments include the nonreciprocal mode conversion (NRMC) optical isolator \cite{hutchingsQuasiPhaseMatchedFaradayRotation2013}, the microring-resonator optical isolator \cite{pintusDesignMagnetoOpticalRing2011,huangElectricallyDrivenThermally2016}, and the Mach-Zehnder interferometer (MZI) optical isolator \cite{shojiSiliconMachZehnder2014, huangIntegratedBroadbandCe2017}. The NRMC isolator utilizes quasi-phase-matched cladding to overcome the modal birefringence of the waveguide. Although it delivers strong isolation performance, the device's size is a few millimeters long, which makes it less ideal for integration. On the other hand, the microring isolator can achieve a strong isolation ratio but faces limitations in narrow isolation bandwidth due to its resonant nature. Lastly, the MZI isolator demonstrates strong performance in terms of isolation ratio, footprint, and isolation bandwidth. However, the device must be fabricated with very high precision to achieve broadband isolation.

An ideal integrated optical isolator should be compact, provide a high isolation ratio in a broad bandwidth, have low insertion loss, and be compatible with large-scale integration. Moreover, it is also essential for the isolator to operate in the TE mode, as most integrated lasers emit light in this mode. The approaches mentioned above have been refined for TE mode, including the NRMC \cite{zhangMonolithicallyIntegratedTEmode1D2017, srinivasanHighGyrotropySeedlayerFreeCe2019}, the microring isolator 
\cite{liuTEmodeMagnetoopticalIsolator2022} and the MZI isolator \cite{yanWaveguideintegratedHighperformanceMagnetooptical2020, yamaguchiLowlossWaveguideOptical2018, zhangMonolithicIntegrationBroadband2019, maIntegratedPolarizationindependentOptical2021}. Since the microring and the MZI isolators are originally TM-based, the TE isolators have been realized by adding a polarization rotator or depositing MO material on the waveguide's sidewall. Despite having a high isolation ratio, these approaches still have narrowband isolation imposed by the previously mentioned constraints.

This paper introduces a novel TE isolator principle based on modal beating in a TMOKE coupled-waveguide system. The idea is to use the TMOKE coupled mode, which is already explored in magneto-plasmonics \cite{abadianBroadbandPlasmonicIsolator2021, hoSwitchablePlasmonicRouters2018}, in classic waveguide configuration. Without the constraints of resonance or MZI interferometers and with the capability to deposit MO material on the waveguide's sidewall, the simulated device demonstrates significant promise in terms of footprint, isolation bandwidth, insertion loss, and scalability. The following section will outline the formalism of the TMOKE couple-waveguide system. We will then discuss how to design an optical isolator based on this system, concluding with an analysis of design parameters and the impact of propagation loss on the device's performance. 

\section{Basic theory of TMOKE coupled-waveguide system}

\begin{figure}[htbp]
    \centering
    \subfloat[]{
        \includegraphics[width=0.8\linewidth]{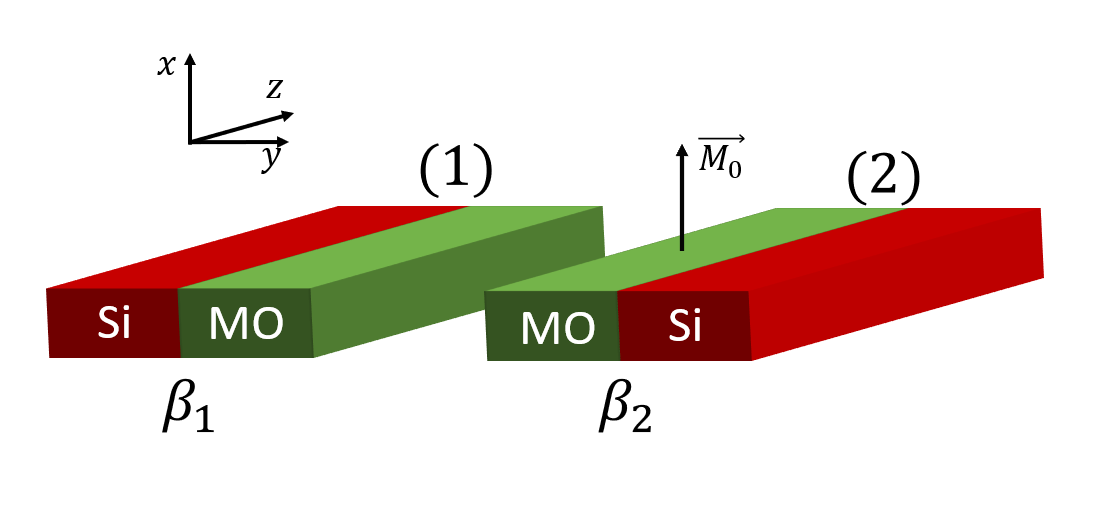}
        \label{fig:bt-tmoke_waveguides}
    }
    \\
    \subfloat[]{
        \includegraphics[width=0.8\linewidth]{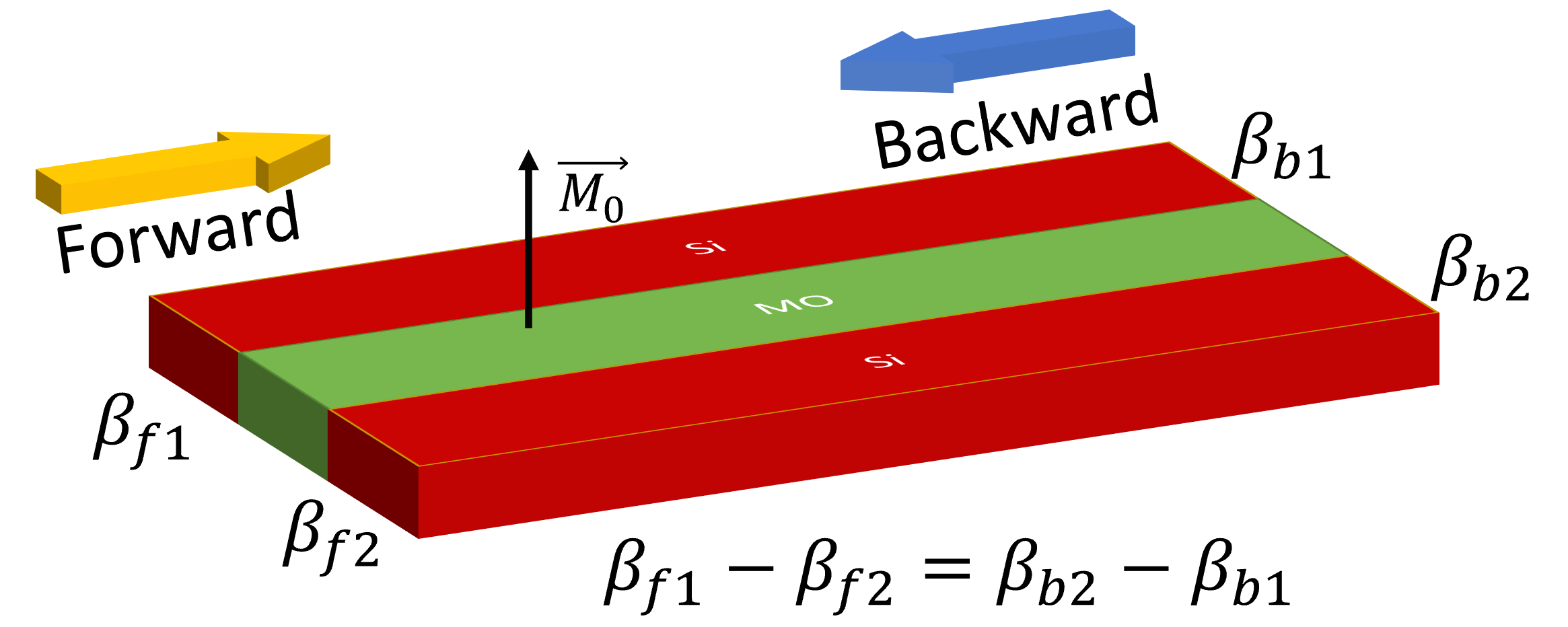}
        \label{fig:bt-tmoke_cws}
    }
    \caption{TMOKE structures. Figure \ref{fig:bt-tmoke_waveguides} depicts a TMOKE waveguide and its mirror image in \(y\)-direction. Figure \ref{fig:bt-tmoke_cws} depicts a TMOKE coupled-waveguide system.}
    \label{fig:bt-tmoke_structures}
\end{figure}

Magneto-optical material breaks the reciprocity of light due to imaginary nondiagonal elements in its permittivity tensor. These imaginary off-diagonal elements lead to numerous magneto-optical effects, including the Faraday and magneto-optical Kerr effects.
TMOKE waveguide, also known as a nonreciprocal phase shifter \cite{zhuromskyyAnalysisPolarizationIndependent1999}, results from interfacing a dielectric waveguide with a MO material, in which the magnetization is perpendicular to the direction of the propagation and the light polarization. There are two main configurations of TMOKE waveguides corresponding to TE and TM modes. This paper shows the TE configuration in Figure \ref{fig:bt-tmoke_waveguides}, where the waveguide's sidewall is attached to the MO material, and the magnetization lies in the \(x\)-direction. In this transversal setup, a pair of nondiagonal elements in the permittivity tensor are activated, resulting in the following form:
\begin{equation}\label{eq:permittivity}
    \mathbf{\epsilon} = \begin{pmatrix}
        \epsilon_0 & 0          & 0          \\
        0          & \epsilon_0 & jg         \\
        0          & -jg        & \epsilon_0
    \end{pmatrix}.
\end{equation}
Here, \(g\) represents the gyrotropy of the MO material. Gyrotropy is usually described by the saturated Faraday rotation  \(\theta_F\), which relates to \(g\) through the equation \(g = 2n\theta_F/k_0\), where \(n\) is the refractive index of the material and \(k_0\) is the wavenumber in the vacuum.

In perturbation theory, as discussed in \cite{zhuromskyyAnalysisPolarizationIndependent1999}, the gyrotropy of the medium induces a shift in the propagation constant of the TMOKE waveguide. In the forward direction, where the optical field is defined by \(\mathbf{E} = \mathbf{E}(x,y) e^{j(\omega t - \beta_f z)}\), the propagation constant is shifted by \(\delta \beta\). It can be represented as \(\beta_f = \beta + \delta \beta\), where \(\beta\) is the propagation constant of the nonpertubative system, and in TE mode case,
\begin{equation}\label{eq:delta_beta}
    \delta \beta = \displaystyle{\frac{2\omega \epsilon_0}{\beta} \frac{\iint g\operatorname{Re}\left[E_y^*\partial_y E_y) \right]\,dxdy}{\iint \left[ \mathbf{E}\times\mathbf{H}^* + \mathbf{E}^*\times\mathbf{H}\right]_zdxdy}.}
\end{equation}
Conversely, the backward propagation field \(\mathbf{E} = \mathbf{E}(x,y) \text{e}^{j(\omega t + \beta_b z)}\) can be considered as a forward propagation with a reversed sign of the gyrotropy. Thus, the propagation constant is given by \(\beta_b = \beta - \delta \beta\). Then, the forward and backward propagation constants of a TMOKE waveguide are shifted by \(2\delta \beta\).

Both the direction of wave propagation and the position of the gyrotropic medium impact the phase shift \(\delta \beta\). When there is a defined direction of wave propagation, structures (1) and (2) in Figure \ref{fig:bt-tmoke_waveguides} show the two possible positions of the MO material within the TMOKE waveguide for the TE mode. Structure (2) is the mirror image of structure (1) flipped in \(y\)-direction. This transformation \(y \longrightarrow -y\) changes the sign of \(\partial_y\) in equation (\ref{eq:delta_beta}), resulting in the reversal of the sign of \(\delta \beta\). As a result, the propagation constants of both structures in both directions are related such that \(\beta_{f1} = \beta_{b2}\), \(\beta_{f2} = \beta_{b1}\) and \(\beta_{f1} - \beta_{f2} = 2\delta \beta\).

When two waveguides are close to each other, they become electromagnetically coupled. This coupling can be synchronous, where the propagation constants of individual waveguides are equal, or asynchronous, where they differ. In the TMOKE coupled-waveguide system depicted in Figure \ref{fig:bt-tmoke_cws}, the gyrotropic medium creates a difference between the propagation constants of the TMOKE waveguide and its mirrored image, as discussed above. This results in asynchronous coupling in the system. To describe the propagation of the asynchronous coupling in the TMOKE coupled-waveguide system, we use the couple mode theory as discussed in \cite{chuang2012physics}.

\begin{figure}[htbp]
    \centering
    \includegraphics[width=0.9\linewidth]{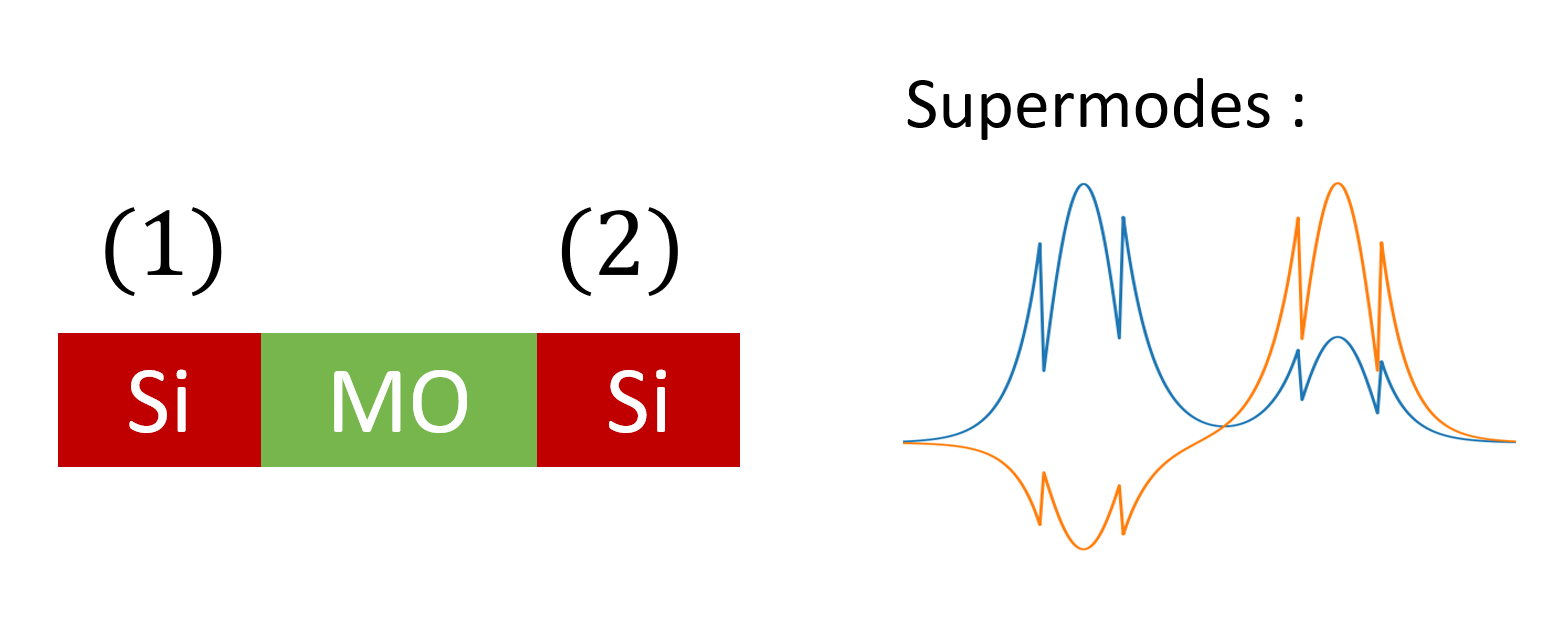}
    \caption{Even and odd supermodes of TMOKE coupled-waveguide system}
    \label{fig:supermodes}
\end{figure}

In coupled mode theory, the total field in a coupled-waveguide system can be decomposed as a linear combination of the supermodes of the system such that
\begin{equation}\label{eq:linear_combination}
    \mathbf{E}(x,y,z) = \Tilde{a}_e\mathbf{E}_e(x,y)\text{e}^{-j\beta_e z} + \Tilde{a}_o\mathbf{E}_o(x,y)\text{e}^{-j\beta_o z} \quad,
\end{equation}
where \(\Tilde{a}_{e,o}\) are the field excitation coefficients, \(\mathbf{E}_{e,o}\) denote the optical fields, and \(\beta_{e,o} \) represent the propagation constants of the even and odd supermodes, respectively. In weak coupling case, the supermodes can be expressed as a linear combination of the two individual waveguide modes, \(\mathbf{E}^{(1)}\) and \(\mathbf{E}^{(2)}\) such that \(\mathbf{E}_{e,o}(x,y) = a_{e,o}\mathbf{E}^{(1)}(x,y) + b_{e,o}\mathbf{E}^{(2)}(x,y)\), where \(a_{e,o}\) and \(b_{e,o}\) denote the amplitude of the supermodes inside the individual waveguides (1) and (2) as shown in Figure \ref{fig:supermodes}, respectively. To simplify the notation, we express these supermodes as normalized column vectors, with their components being the amplitude \(a_{e,o}\) and \(b_{e,o}\) \cite{yariv1973coupled}, \(\mathbf{E}_{e,o} = \displaystyle{\frac{1}{\sqrt{a_{e,o}^2 + b_{e,o}^2}}} \begin{pmatrix} a_{e,o} \\ b_{e,o} \end{pmatrix}\). In asynchronous coupling where \(\beta_1 - \beta_2 > 0\), the amplitudes of the supermodes \(a_{e,o}\) and \(b_{e,o}\) are related such that \(\displaystyle{\frac{b_e}{a_e}} = -\frac{a_o}{b_o} = C\), where \(0 < C \leq 1\). \(C\) then denotes the symmetry ratio of the supermodes.

In the TMOKE coupled-waveguide system, if the forward propagation satisfies the condition \(\beta_{f1} - \beta_{f2} > 0\), then the even and odd supermodes in the forward direction, \(\mathbf{E}_{fe}\) and  \(\mathbf{E}_{fo}\),  can be expressed as:
\begin{equation}\label{eq:supermodes_forward}
    \mathbf{E}_{fe} = \frac{1}{\sqrt{1 + C^2}} \begin{pmatrix} 1 \\ C \end{pmatrix}, \quad \mathbf{E}_{fo} = \frac{1}{\sqrt{1 + C^2}} \begin{pmatrix} -C \\ 1 \end{pmatrix}.
\end{equation}
On the other hand, in the backward direction, the propagation constants are swapped such that  \(\beta_{b1} - \beta_{b2} = \beta_{f2} - \beta_{f1} < 0\). Thus, the amplitudes of both even and odd supermodes in the backward direction, \(\mathbf{E}_{be}\) and  \(\mathbf{E}_{bo}\),  are also swapped:
\begin{equation}\label{eq:supermodes_backward}
    \mathbf{E}_{be} = \frac{1}{\sqrt{1 + C^2}} \begin{pmatrix} C \\ 1 \end{pmatrix}, \quad \mathbf{E}_{bo} = \frac{1}{\sqrt{1 + C^2}} \begin{pmatrix} -1 \\ C \end{pmatrix}.
\end{equation}
The propagation constant of the supermodes \(\beta_{e,o}\) remains unchanged, even though the amplitudes are interchanged. The interchanged amplitudes in the TMOKE coupled-waveguide system distinguish it from the typical asynchronous coupled-waveguide system. This results in nonreciprocal propagation within the system and is crucial in designing the optical isolator, which is discussed in the following section.

In lossless case, the propagation field is expressed as in equation (\ref{eq:linear_combination}), where the excitation coefficients \(\Tilde{a}_{e,o}\) are determined by the initial injection \(\mathbf{E}_{in} = \frac{1}{\sqrt{a_{in}^2 + b_{in}^2}} \begin{pmatrix} a_{in} \\ b_{in} \end{pmatrix}\). Knowing that the coupling is asynchronous and the supermodes of the system are expressed in equations (\ref{eq:supermodes_forward}) and (\ref{eq:supermodes_backward}), the propagating field in both directions at a distance \(z\) can be described by \(2\times2\) transfer matrices such that, \(\mathbf{E}_{f,b \; out} = \text{e}^{\pm j\beta_e z} \mathbf{T}_{f,b}(z)\mathbf{E}_{f,b \; in} \). The transfer matrices for forward and backward propagations are given by:
\begin{equation}\label{eq:transfer_forward}
    \mathbf{T}_f(z) = \displaystyle{\frac{1}{1+C^2}}\begin{pmatrix}
        1 + C^2 \text{e}^{j2\psi z}         & C\left(1-\text{e}^{j2\psi z}\right) \\
        C\left(1-\text{e}^{j2\psi z}\right) & C^2 +  \text{e}^{j2\psi z}
    \end{pmatrix},
\end{equation}

\begin{equation}\label{eq:transfer_backward}
    \mathbf{T}_b(z) = \displaystyle{\frac{1}{1+C^2}}\begin{pmatrix}
        C^2 +  \text{e}^{-j2\psi z}          & C\left(1-\text{e}^{-j2\psi z}\right) \\
        C\left(1-\text{e}^{-j2\psi z}\right) & 1+  C^2\text{e}^{-j2\psi z}
    \end{pmatrix}
\end{equation}
respectively, where \(\psi = \displaystyle{\frac{\beta_e - \beta_o}{2}}\). Despite the reversal sign of \(z\), the diagonal elements of the backward transfer matrix are interchanged compared to the forward transfer matrix. This happens because the amplitudes of the supermodes are interchanged when the direction of propagation is inverted. This demonstrates the nonreciprocity of the system, where \( \mathbf{T}_f^\mathrm{T} \neq \mathbf{T}_b \) \cite{yamamoto1974circuit}. In contrast to a normal coupled system, another distinguishing feature of the TMOKE coupled-waveguide system is the ability to modify the output field without changing the input. In addition to the input, the output field can be adjusted by tuning the ratio \(C\) and \(\psi\) of the supermodes, which are determined by the magneto-optical effect and the coupling strength of the system.

The outputs of synchronous and asynchronous coupled-waveguide systems exhibit a common behavior known as self-imaging, regardless of the magneto-optical effect. These coupled systems will reproduce the input at a propagation distance of a beating length, which is defined by \(L = \displaystyle{\frac{2\pi}{\beta_e - \beta_o}}\). In the TMOKE case, both transfer matrices presented in equations (\ref{eq:transfer_forward}) and (\ref{eq:transfer_backward}), \(\mathbf{T}_f(z)\) and \(\mathbf{T}_b(z)\), become identity matrices when \(z = L\). As a result, \(\mathbf{E}_{f, b \;out} \) and \( \mathbf{E}_{f,b\; in}\) are the same regardless of the phase term.

\begin{figure}[htbp]
    \centering
    \subfloat[]{
        \includegraphics[width=0.8\linewidth]{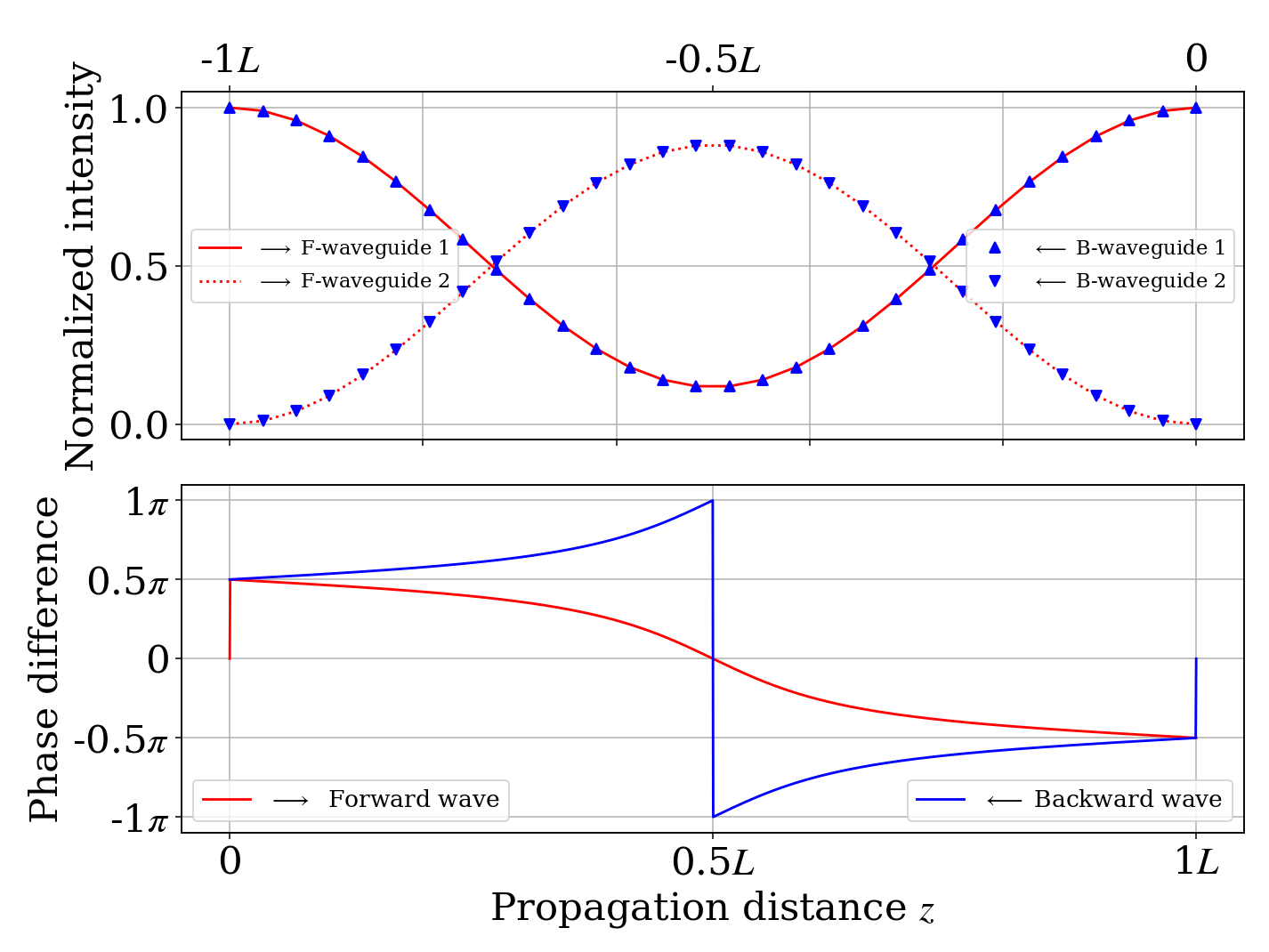}
        \label{fig:intensity_plot_tmoke_cws_c07}
    }
    \\
    \subfloat[]{
        \includegraphics[width=0.8\linewidth]{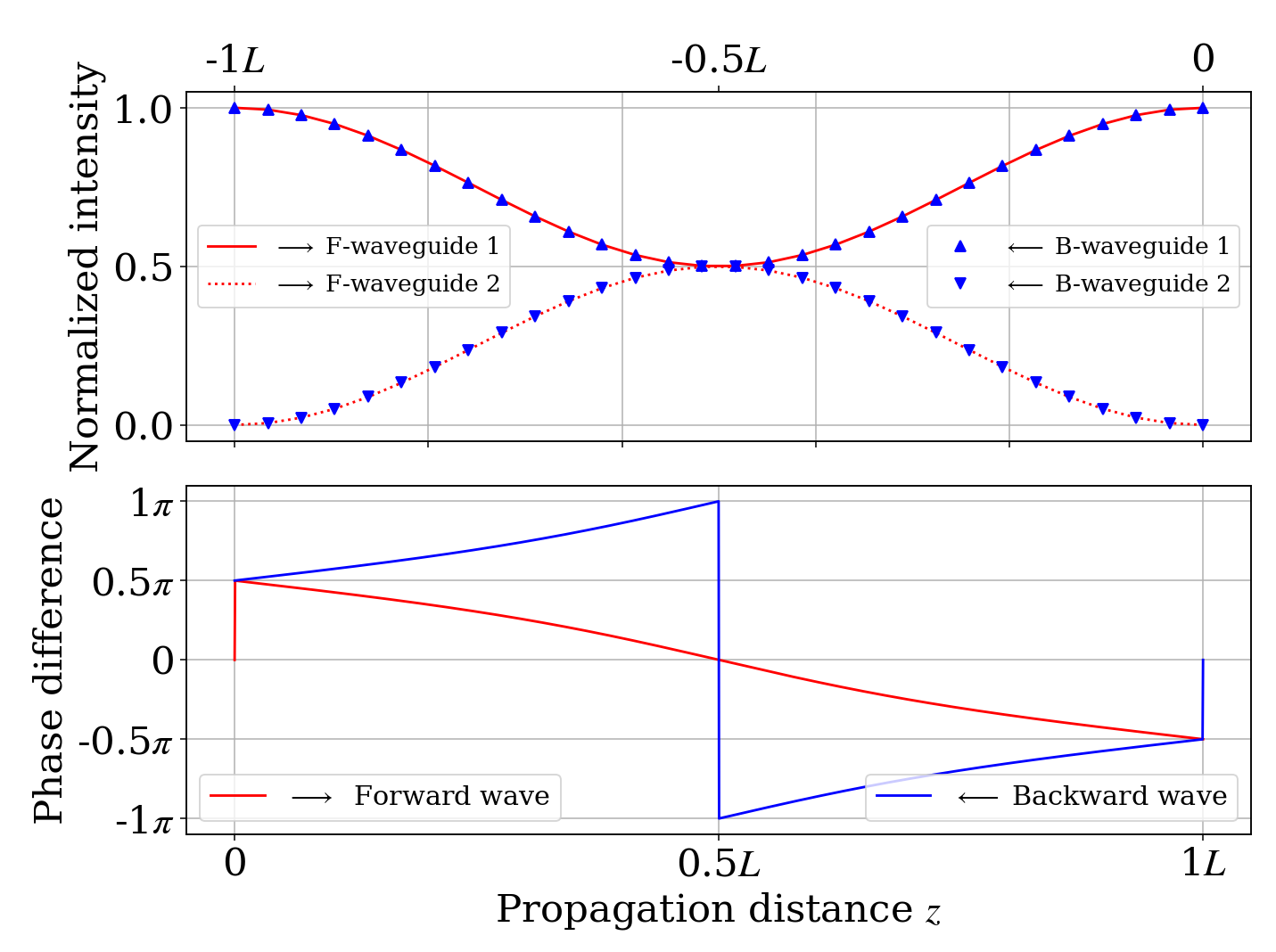}
        \label{fig:intensity_plot_tmoke_cws_c_0414}
    }
    \caption{Optical intensity \((\|a_{f,b \;out}\|^2,\|b_{f,b \;out}\|^2)^\mathrm{T} \) and phase difference \(\mathrm{arg}(a_{f,b \;out} - b_{f,b \;out})\) in TMOKE coupled-waveguide system when injecting \(\mathbf{E}_{in}=(1,0)^\mathrm{T} \) in both forward and backward direction. The forward and backward propagation correspond to positive and negative propagation distances, respectively. Figure \ref{fig:intensity_plot_tmoke_cws_c07} and \ref{fig:intensity_plot_tmoke_cws_c_0414} are for \(C = 0.7\) and \(\sqrt{2}-1\), respectively.}
    \label{fig:intensity_plot_tmoke_cws}
\end{figure}

For any \(z\), the output optical field \((a_{out}, b_{out})^\mathrm{T}\) of any input field \((a_{in}, b_{in})^\mathrm{T}\) can be decomposed as a linear combination of the basis outputs such that: \(\mathbf{E}_{f,b\;out} = a_{in} \mathbf{T}_{f,b}(1,0)^\mathrm{T} + b_{in} \mathbf{T}_{f,b}(0,1)^\mathrm{T}\). Figure \ref{fig:intensity_plot_tmoke_cws} depicts the basic output \(\mathbf{T}_{f,b}(z)(1,0)^\mathrm{T}\) in function of \(z\). At a propagation distance of \(z = \frac{L}{2}\), the system also exhibits peculiar behavior in addition to self-imaging at \(z=L\). The transfer matrices at distance \(z = L/2\), \(\mathbf{T}_f(L/2)\) and \(\mathbf{T}_b(L/2)\), are respectively given by:
\[\mathbf{T}_f(L/2) = \displaystyle{\frac{1}{1+C^2}}\begin{pmatrix}
        1-C^2 & 2C     \\
        2C    & C^2 -1
    \end{pmatrix},\]

\[\mathbf{T}_b(L/2) = \displaystyle{\frac{1}{1+C^2}}\begin{pmatrix}
        C^2 -1 & 2C    \\
        2C     & 1-C^2
    \end{pmatrix}.\]
The diagonal elements of these matrices have opposite signs. Regardless of the ratio \(C\) as depicted in Figure \ref{fig:intensity_plot_tmoke_cws_c07} and \ref{fig:intensity_plot_tmoke_cws_c_0414}, the opposite signs result in a \(\pi\)-phase shift between the phase difference of backward and forward basis output pairs: \([\,\mathbf{T}_{f}(1,0)^\mathrm{T},\mathbf{T}_{b}(1,0)^\mathrm{T}]\,\) and \([\,\mathbf{T}_{f}(0,1)^\mathrm{T}, \mathbf{T}_{b}(0,1)^\mathrm{T}]\,\), where the phase difference is defined by \(\mathrm{arg}(a_{out}-b_{out}) \).

In the case where \(C = \sqrt{2} - 1\) and the distance \(z = L/2\), the forward and backward optical inputs, \((1,0)_{f,b}^\mathrm{T}\) and \((0,1)_{f,b}^\mathrm{T}\), are split into two equal-intensity beams in addition to the \(\pi\)-phase shift. Similarly, the two equal-intensity beams, \(\frac{1}{\sqrt{2}}(1,1)^\mathrm{T}\) and \(\frac{1}{\sqrt{2}}(-1,1)^\mathrm{T}\), are recombined after \(L/2\) propagation. A detailed mapping between these inputs and outputs is presented in Table (\ref{tab:input-output}).

\bgroup
\def\arraystretch{1.5}
\begin{table}[htbp]
\begin{ruledtabular}
    \begin{tabular}{ccc}
    Input                                   & Forward output L/2                      & Backward output L/2                     \\
    \hline
    \((1,0)^\mathrm{T}\)                    & \(\frac{1}{\sqrt{2}}(1,1)^\mathrm{T}\)  & \(\frac{1}{\sqrt{2}}(-1,1)^\mathrm{T}\) \\
    \((0,1)^\mathrm{T}\)                    & \(\frac{1}{\sqrt{2}}(-1,1)^\mathrm{T}\) & \(\frac{1}{\sqrt{2}}(1,1)^\mathrm{T}\)  \\
    \(\frac{1}{\sqrt{2}}(1,1)^\mathrm{T}\)  & \((1,0)^\mathrm{T}\)                    & \((0,1)^\mathrm{T}\)                    \\
    \(\frac{1}{\sqrt{2}}(-1,1)^\mathrm{T}\) & \((0,1)^\mathrm{T}\)                    & \((1,0)^\mathrm{T}\)                    \\
\end{tabular}
\end{ruledtabular}

\caption{Summary of the input-output mapping in case where \(C = \sqrt{2}-1\) and \(z = L/2\)}
\label{tab:input-output}
\end{table}
\egroup

The TMOKE coupled-waveguide system combines two effects: the nonreciprocal phase shift in the TMOKE waveguide and the modal beating in the coupled-waveguide system. By adjusting \(C=\sqrt{2} - 1\) and \(z = L/2\) appropriately, the system shows the splitting and recombination of light, along with a \(\pi\)-phase shift as outlined in Table (\ref{tab:input-output}). This input-output mapping is important for designing the optical isolator, which will be discussed in the next section.

\section{TMOKE couple-waveguide optical isolator}
\subsection{Operation principle}
\begin{figure}[ht]
    \centering
    \includegraphics[width=\linewidth]{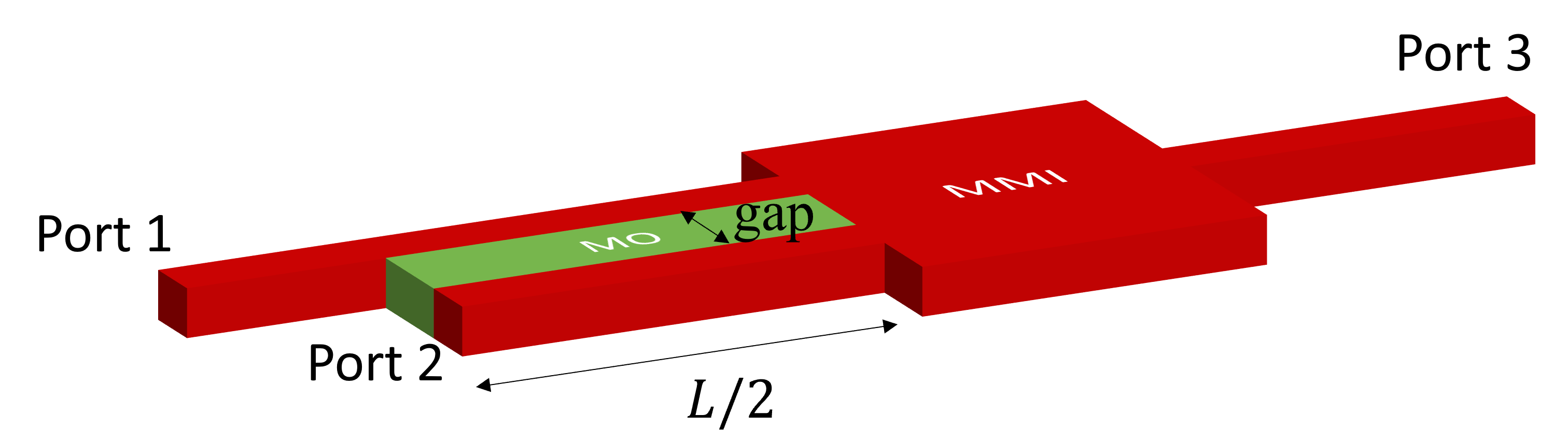}
    \caption{Schematic of the optical isolator, a combination of a TMOKE coupled-waveguide system and a \(1\times2\) multimode interferometer.}
    \label{fig:schema_isolator}
\end{figure}

The TMOKE coupled-waveguide system is nonreciprocal and asynchronous due to the perturbation caused by the gyrotropic medium. When properly calibrated, it shows useful behaviors such as beam splitting, recombination, and a \(\pi\)-phase shift, as detailed in Table \ref{tab:input-output}. This part introduces a new design for an optical isolator that exploits these properties. The structure of the optical isolator, as depicted in Figure \ref{fig:schema_isolator}, consists of a well-calibrated TMOKE coupled-waveguide system and a \(1\times2\) multimode interferometer (MMI). The structure has three ports. Ports 1 and 2 are associated with the TMOKE coupled-waveguide system and are connected to the two ports of the MMI. Port 3 is linked to the center port of the MMI. 

To describe the operation principle of the optical isolator, we use the formalism presented in the previous section. For consistency with the last section, we consider the lossless case and the configuration of the TMOKE coupled-waveguide structure where the forward propagation constant of the waveguide associated with port 1 is greater than that of port 2 (\(\beta_{f1} - \beta_{f2} > 0\)). The output and input optical fields of port 1 and port 2 are associated and presented in amplitude-column vector \((a, b)^\mathrm{T}\), where \(a\) and \(b\) correspond to the amplitude of the optical field in port 1 and 2, respectively. Considering the forward propagation is from port 1 to port 3, the output can be written as the product of matrices such that :

\begin{equation} \label{eq:output_oi_f}
    \mathbf{E}_{f,\;out} = \mathbf{T}_{r, \text{MMI}}\mathbf{T}_f\mathbf{E}_{f,\;in},
\end{equation}
where \(\mathbf{E}_{f,in}\) represents the input in port 1, given by \(\mathbf{E}_{f,\;in} = (1,0)^\mathrm{T}\) and \(\mathbf{T}_{r, \text{MMI}} = \frac{1}{\sqrt{2}}(1\quad 1)\) denotes the recombining transfer matrix of the MMI. Additionally, \(\mathbf{T}_f\) denotes the transfer matrix of the calibrated TMOKE coupled-waveguide system (\(C = \sqrt{2}-1\) and \(z = L/2\)), given by:

\begin{equation}\label{eq:transfer_matrix_calib_f}
    \mathbf{T}_f = \displaystyle{\frac{1}{\sqrt{2}}}\begin{pmatrix}
        1 & 1  \\
        1 & -1
    \end{pmatrix}.
\end{equation}
This product gives a scalar output, \(\mathbf{E}_{f,\;out}\) = 1, which also means that the output at port 3, \(\mathbf{E}_{f,\;out}\), is fully transmitted when injecting light at port 1.

Conversely, the backward propagation from port 3 to port 1 is described by
\begin{equation}\label{eq:output_oi_b}
    \mathbf{E}_{b,\;out} = \mathbf{T}_b\mathbf{T}_{s, \text{MMI}}\mathbf{E}_{b,\;in},
\end{equation}
where \(\mathbf{E}_{b,\;in}\) represents the input in port 3 and is denoted by a scalar value 1, \(\mathbf{T}_{s, \text{MMI}} = \frac{1}{\sqrt{2}}(1,1)^\mathrm{T}\) is the splitting transfer matrix of the MMI and \(\mathbf{T}_b\) is the backward calibrated transfer matrix of TMOKE coupled waveguide system, given by:
\begin{equation}\label{eq:transfer_matrix_calib_b}
    \mathbf{T}_b = \displaystyle{\frac{1}{\sqrt{2}}}\begin{pmatrix}
        -1 & 1 \\
        1  & 1
    \end{pmatrix}.
\end{equation}
The product gives an output of \(\mathbf{E}_{b,\;out} = (0,1)^\mathrm{T}\), where 0 and 1 are the output at port 1 and port 2, respectively. This indicates that the backward output of the optical isolator (the output at port 1) is fully blocked, and the input is transmitted to port 2 instead.

The device can also act as an optical isolator when port 2 and port 3 are used as the input-output pair. Port 2 denotes the lower forward-propagation-constant waveguide in the TMOKE coupled-waveguide system. With this choice of input-output pair, the device must operate in the opposite direction to the one described previously to function as an optical isolator. When operating in three-to-two direction, the output is given by \( \mathbf{E}_{out} = \mathbf{T}_b \mathbf{T}_{s,\text{MMI}}\mathbf{E}_{in} \). This product yields the output of \( (0,1)^\mathrm{T} \), indicating that the light is fully transmitted through port 2. In the opposite direction, from port 2 to port 3, the output is expressed as \( \mathbf{E}_{out} = \mathbf{T}_{r,\text{MMI}}\mathbf{T}_f \mathbf{E}_{in} \), resulting in an output of 0 at port 3 due to destructive interference within the MMI. Hence, the device achieves the function of an optical isolator in the same configuration but with a different input-output pair and operation direction.

There are thus two possible configurations to use the design proposed in Figure \ref{fig:schema_isolator}: the one-to-three optical isolator and the three-to-two optical isolator. One downside of the three-to-two optical isolator is the occurrence of destructive interference in the MMI. It is important to note that the destructive interference in MMI does not eliminate the optical field; instead, it converts the optical field to a radiative mode. Besides, the one-to-three optical isolator allows the collection of the feedback signal at port 2, which offers an additional function to the device. For these reasons, the one-to-three optical isolator appears to be the best design, and the subsequent discussion will mainly focus on it.

\subsection{Design and performance}

The optical isolator, as depicted in Figure \ref{fig:schema_isolator}, consists of two main parts: a TMOKE system and a \(1\times2\) MMI. The design of the MMI part is straightforward as it is intended to align its output with the waveguides' separation of the TMOKE system \cite{soldanoOpticalMultimodeInterference1995}. Therefore, the primary focus of the isolation design is calibrating the TMOKE coupled-waveguide system. The system must be calibrated for \(C = \sqrt{2} - 1\) to achieve maximum optical isolation. The ratio \(C\) is determined through the supermodes, which are decomposed into amplitudes of two individual waveguides. 

In the TMOKE coupled-waveguide system where \(\beta_1 - \beta_2 > 0\), the ratio \(C\) is consistent for both supermodes and is given by \(C = \displaystyle{\frac{\psi-\Delta}{K}}\). Here, \(K\) represents the coupling strength of the coupled system, \(\Delta\) is the difference in the propagation constants of the individual waveguides (\(\beta_1 - \beta_2\)), and \(\psi\) is the difference in the propagation constants of the supermodes, defined as \(\psi = \frac{\beta_e - \beta_o }{2}= \sqrt{\Delta^2 + K^2}\) \cite{chuang2012physics}. The expression for \(C\) can be simplified to \(\frac{\sqrt{\Delta^2 + K^2} - \Delta}{K}\). For a fixed coupling strength \(K\), the ratio \(C\) decreases as the parameter \(\Delta\) increases, while for a fixed value of \(\Delta\), \(C\) increases as \(K\) increases. This relationship highlights the balance between coupling strength \(K\) and the difference in propagation constants  \(\Delta\) within the TMOKE coupled-waveguide system. Understanding this balance is crucial for optimizing the isolation performance, as it directly affects the supermodes' behavior, i.e., the ratio \(C\).

The difference in propagation constants of individual waveguides \(\Delta\) and the coupling strength \(K\) have distinct origins. The difference \(\Delta\) is expressed as \(\Delta = \beta_1 - \beta_2 = 2\delta\beta\), where \(\delta\beta\) represents the shift caused by the magneto-optical effect. On the other hand, the coupling strength \(K\) depends on the overlap between the evanescent tails of both waveguides' modes. Therefore, the main factor affecting the coupling strength is the separation between the waveguides. However, \(K\) and \(\Delta\) can also be influenced by other parameters, including the geometry of the waveguides and the refractive index of the waveguide and the magneto-optical material. These parameters impact the confinement of the optical field, thereby affecting the values of \(K\) and \(\Delta\). Hence, to calibrate the TMOKE system for a desired value of \(C\), we need to consider adjusting the gyrotropy of the MO material, the separation between the coupled waveguides, the geometry of the waveguides and the refractive indices of the materials.

The design parameters need to meet a specific value of \(C\) while also optimizing the footprint and operational bandwidth of the optical isolator. However, some parameters are limited by the choice of materials. For example, in this design, the silicon-on-insulator (SOI) platform is chosen to guide the optical field, which determines the refractive index and the waveguide geometry. When it comes to MO material, only a limited number have been well-developed and used with the integrated platform \cite{srinivasanReviewIntegratedMagnetooptical2022}. We have selected \ce{Ce_{1.4}Y_{1.6}Fe5O12} which exhibits the highest saturated Faraday rotation of \qty{-5900}{\degree\per\centi\meter} at \qty{1550}{\nano\meter} wavelength in the litterature. This corresponds to \(g = 0.0118\) \cite{yanWaveguideintegratedHighperformanceMagnetooptical2020}. With these material limitations, the separation between the coupled waveguides becomes the sole remaining design parameter to adjust the TMOKE coupled-waveguide system.

\begin{figure}[ht]
    \centering
    \includegraphics[width=0.8\linewidth]{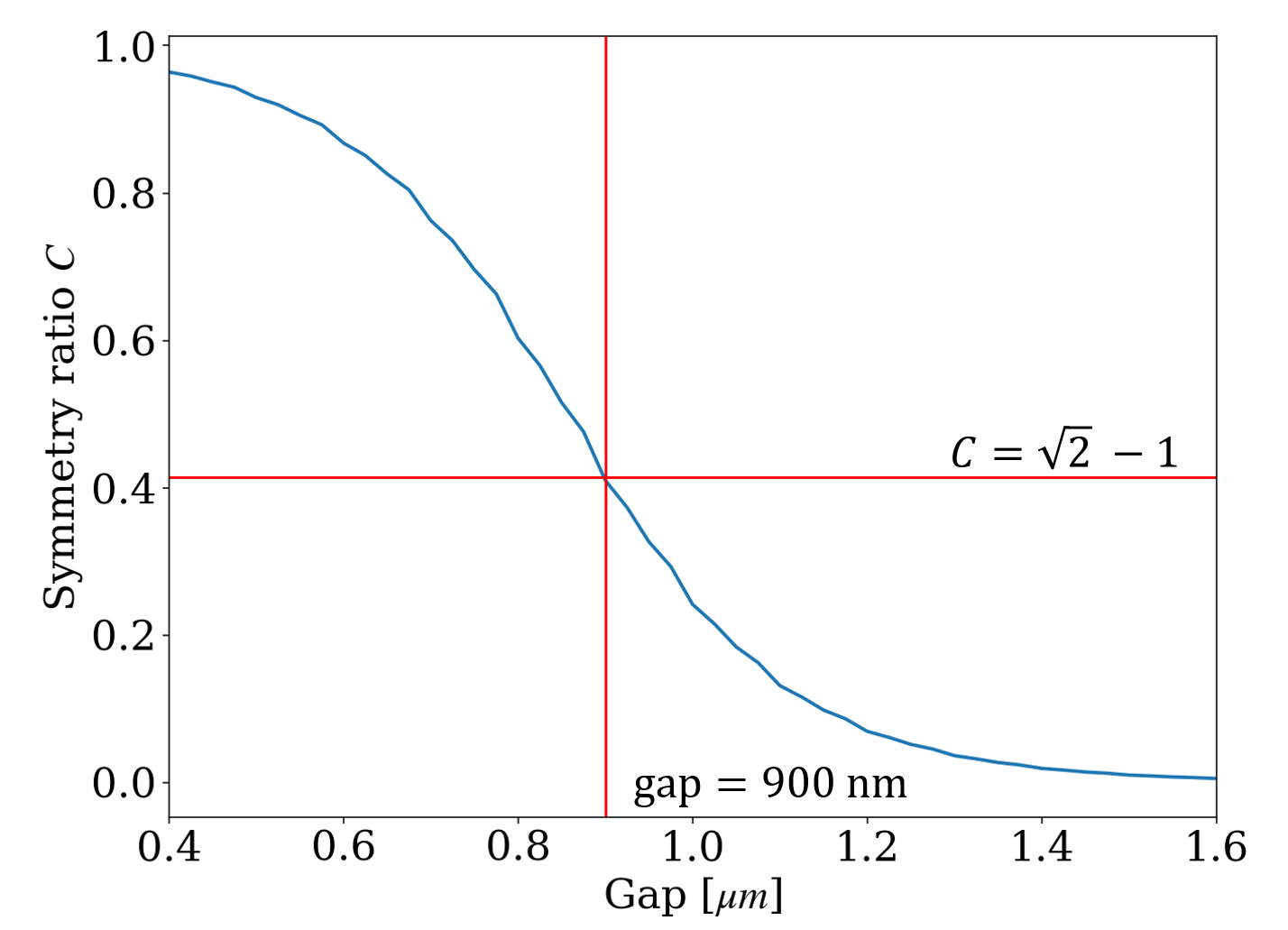}
    \caption{Supermodes' symmetry ratio in function of gap at \(\lambda = \qty{1.55}{\micro\meter}\), \(h=\qty{200}{\nano\meter}\) and \(w=\qty{400}{\nano\meter}\) in SOI platform with \ce{SiO2} top cladding. Gap refers to the separation between both waveguides.}
    \label{fig:pta_vs_gap}
\end{figure}

\begin{figure}[ht]
    \centering
    \includegraphics[width=\linewidth]{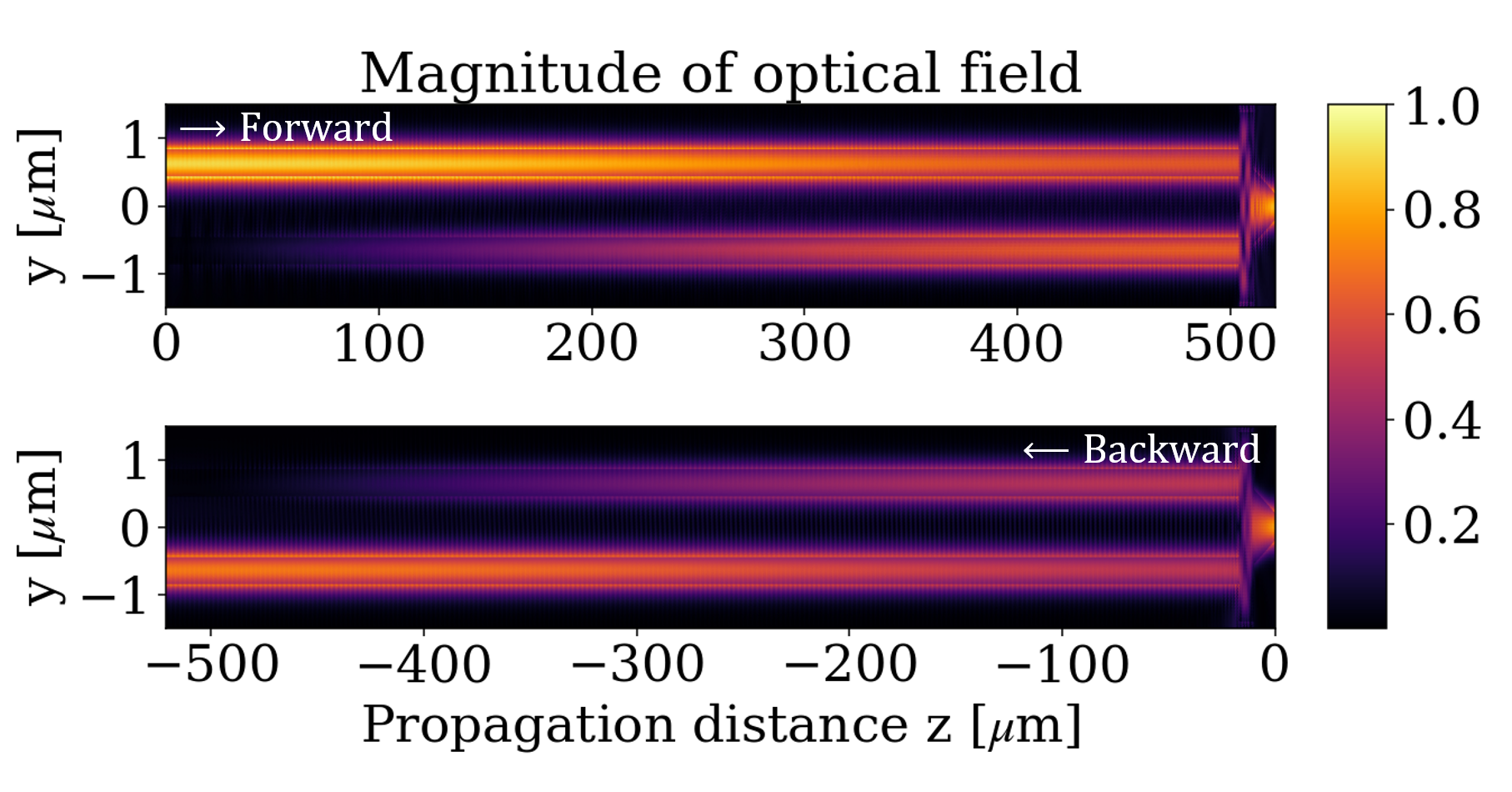}
    \caption{Optical field propagation simulated by 2D FDTD inside the one-to-three optical isolator structure depicted in figure \ref{fig:schema_isolator}. Forward and backward propagation are distinguished by the sign of propagation distance.}
    \label{fig:field_dist_oi}
\end{figure}

Figure \ref{fig:pta_vs_gap} depicts the symmetry ratio \(C\) as a function of the separation between coupled waveguides. In this calculation, the waveguide platform is the SOI platform (which has silica as top cladding), the MO material is \ce{Ce_{1.4}Y_{1.6}Fe5O12}, and the cross section of the waveguide is \qtyproduct{400 x 220}{nm}. To calibrate the TMOKE coupled-waveguide system for the ratio \(C = \sqrt{2} -1 \) at \qty{1550}{nm}, the gap between the coupled waveguides is \qty{900}{nm}. The effective index of the even and odd supermodes are \(n_e = \num{2.3605}\) and \(n_o = \num{2.3590}\), respectively. Therefore, the beating length \(L\) is approximately \qty{1000}{\um}. Figure \ref{fig:field_dist_oi} shows the field distribution simulated by 2D-FDTD for the forward and backward propagation inside the optical isolator using the mentioned configuration. 

\begin{figure}[ht]
    \centering
    \includegraphics[width=0.8\linewidth]{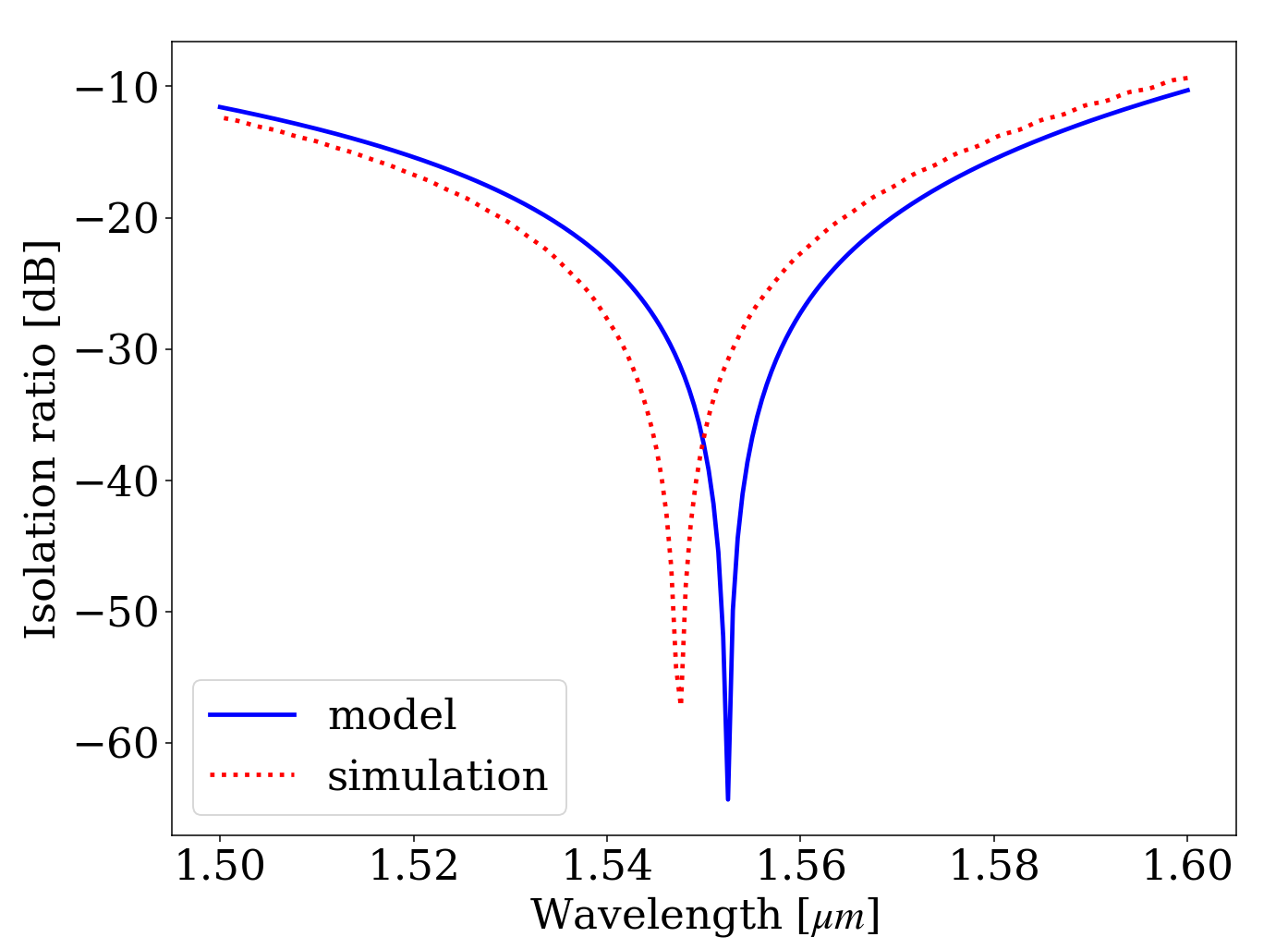}
    \caption{Isolation ratio of the designed optical isolator predicted by the model and FDTD simulation.}
    \label{fig:isolation_ratio}
\end{figure}

When evaluating the performance of the optical isolator, we are interested in two main properties: the maximum isolation ratio and the isolation bandwidth. The isolation ratio is defined by \(10\log{\left(\frac{\text{P}_{\text{backward}}}{\text{P}_{\text{forward}}}\right)}\), where \(\text{P}\) is the output optical intensity of the isolator. The isolation bandwidth can be simulated using FDTD or calculated using the model introduced in the previous section. When we don't take into account the loss and the dispersion of the materials, the wavelength-dependent forward and backward outputs can be expressed respectively as:
\begin{equation}\label{eq:E_f_lambda}
    \mathbf{E}_{f,\;out} = \displaystyle{\frac{1}{\sqrt{2}\left(1+C_\lambda^2\right)}}\left[1 + C_\lambda + C_\lambda\left(C_\lambda -1\right) \text{e}^{j\pi\frac{L_0}{L_\lambda}} \right],
\end{equation}
\begin{equation}\label{eq:E_b_lambda}
    \mathbf{E}_{b,\;out} = \displaystyle{\frac{1}{\sqrt{2}\left(1+C_\lambda^2\right)}}\left[C_\lambda\left(C_\lambda +1\right) + \left(C_\lambda -1 \right)\text{e}^{j\pi\frac{L_0}{L_\lambda}} \right],
\end{equation}
where \(L_0\) is the beating length at \qty{1550}{nm}, \(L_\lambda\) is the beating length at an arbitary wavelength \(\lambda\), \(C_\lambda\) is the ratio \(C\) of the supermodes calculated at an arbitrary wavelength \(\lambda\). Figure \ref{fig:isolation_ratio} shows the comparison of the isolation ratio between the model and the FDTD simulation for the wavelengths of \qtyrange{1.5}{1.6}{\um}. The isolation-ratio curve from the FDTD simulation is shifted from that of the model by \qty{5}{nm}. This \qty{5}{nm} shift results from numerical errors produced by eigenmode calculation and the FDTD simulation. Overall, the design achieves a maximum isolation ratio exceeding \qty{-50}{dB}. Additionally, without taking the material dispersion into account, the \qty{20}{dB}-isolation bandwidth spans approximately \qty{35}{nm}.

\section{Discussion}
\subsection{On the cross section of the waveguide}

\begin{figure}[htbp]
    \centering
    \subfloat[]{
        \includegraphics[width=0.8\linewidth]{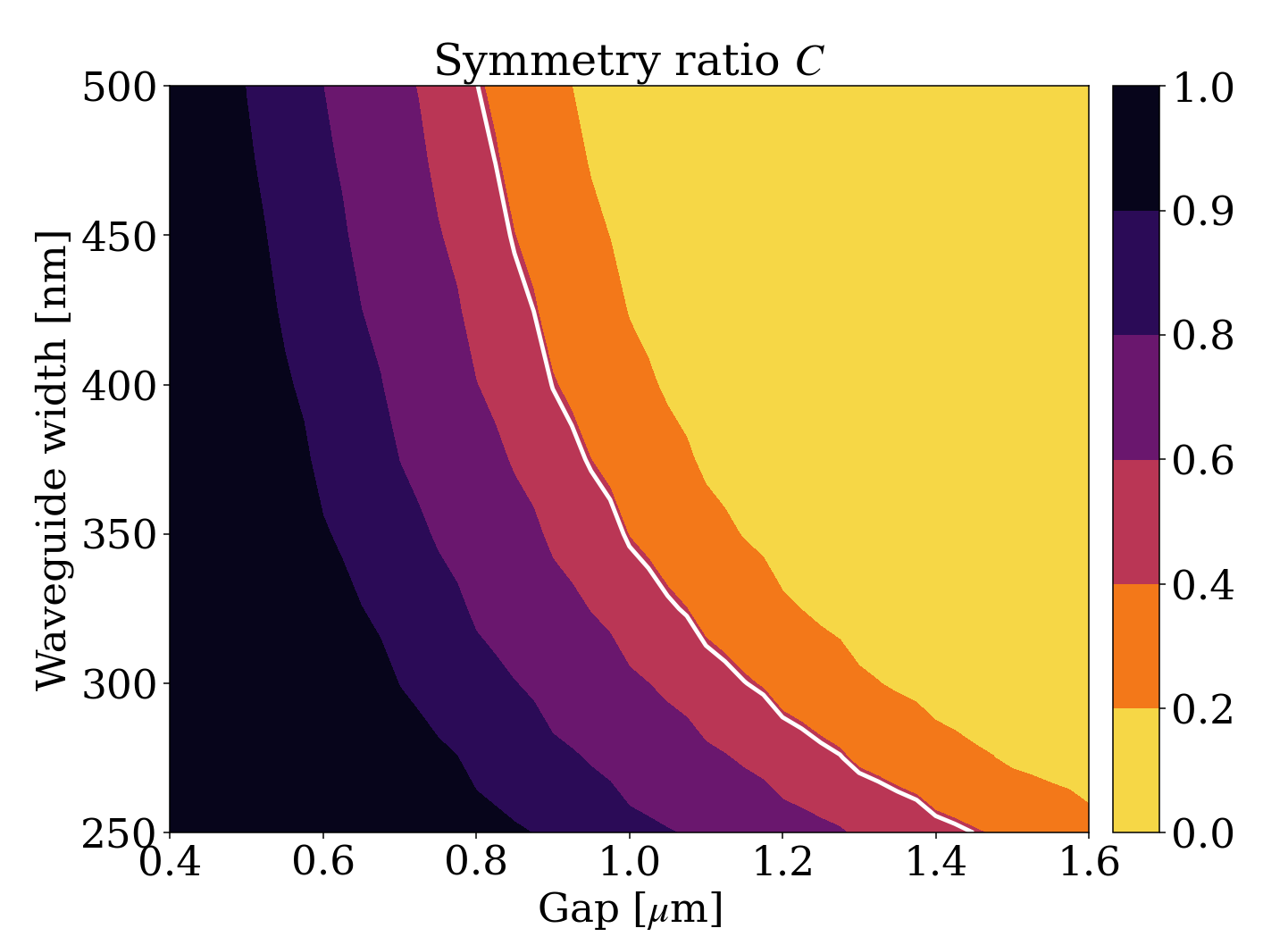}
        \label{fig:w_sweep_C}
    }
    \\
    \subfloat[]{
        \includegraphics[width=0.8\linewidth]{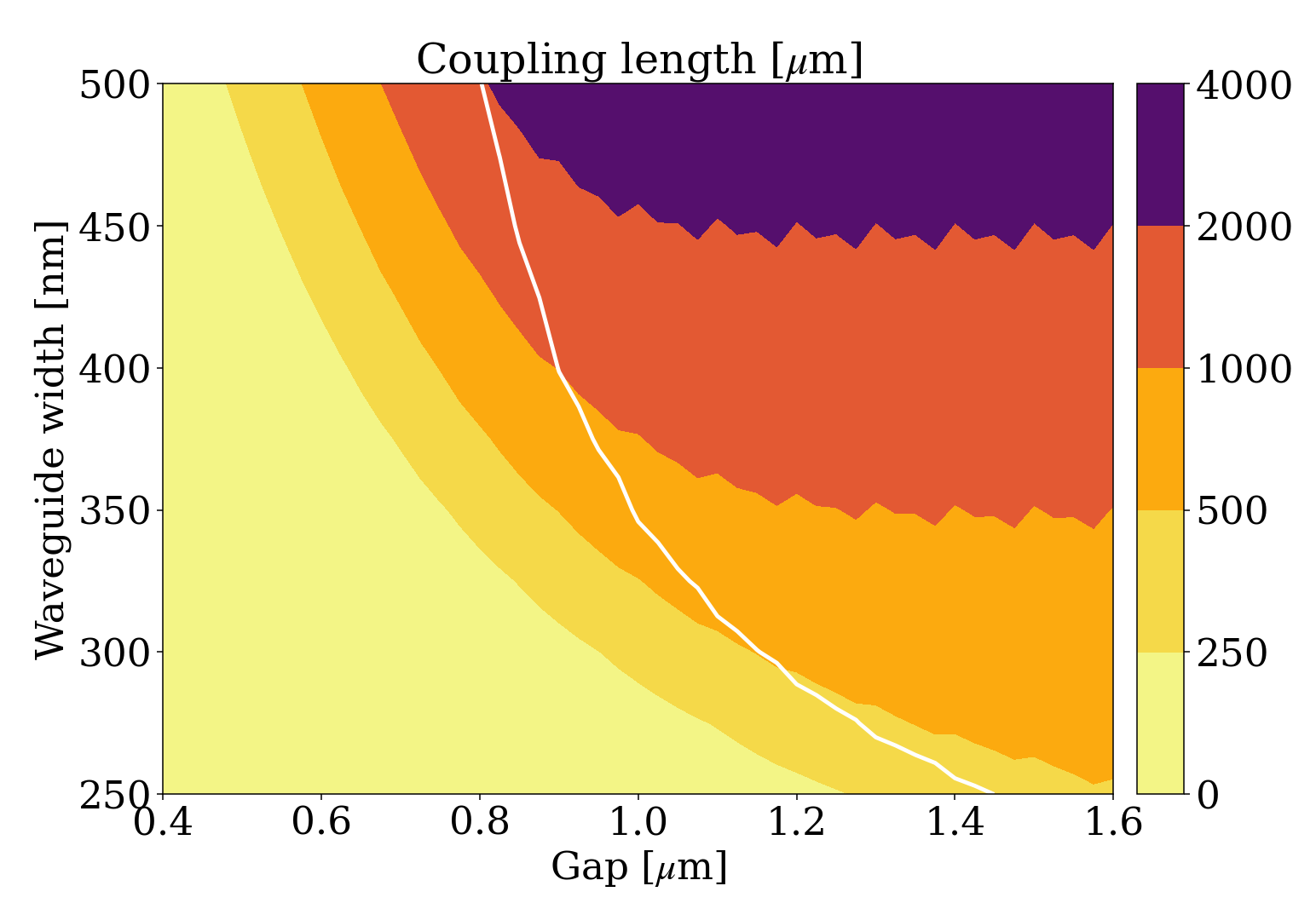}
        \label{fig:w_sweep_L}
    }
    \caption{Supermodes' symmetry ratio (\ref{fig:w_sweep_C}) and beating length (\ref{fig:w_sweep_L}) as a function of gap and waveguide's width.}
    \label{fig:width_sweep}
\end{figure}
To ensure a single-mode propagation in the design presented above, the waveguide's cross section is set to \qtyproduct{400 x 220}{nm}. However, waveguide geometry will directly impact the beating length of the coupled waveguide, i.e., the footprint of the optical isolator. The white line in  Figure \ref{fig:width_sweep} shows a set of configurations (waveguide widths and gaps) where the symmetry ratio \(C\) is equal to \(\sqrt{2}-1\). In each configuration, we can observe the corresponding beating length of the TMOKE system, i.e., the footprint of the optical isolator. Figure \ref{fig:width_sweep} shows that as the waveguide width decreases, the beating length also decreases.

When the width of the waveguide is reduced, the optical isolator has a smaller footprint, but it also experiences more losses. A narrower waveguide means less confinement of the optical field within the waveguide. This leads to higher propagation loss as more of the optical field spreads into the MO material. In summary, given a fixed choice of materials, optimizing the isolation performance in terms of insertion losses would involve balancing the footprint and propagation loss with the waveguide geometry.

\subsection{On the loss of the material}

The optical isolator presented above is in a lossless case. However, the MO material is well-known for lossy propagation. The propagation loss of a TMOKE waveguide reported in \cite{yanWaveguideintegratedHighperformanceMagnetooptical2020} is about \qty{30}{dB\per cm}. To further analyze the robustness of the optical isolator design in the lossy case, we introduce the imaginary part of the propagation constant into the model, \(\Tilde{\beta}_{e,o} = \beta_{e,o} - i\alpha_{e,o}\). By substituting this complex propagation constant into the existing model, the transfer matrix of the TMOKE coupled-waveguide system at \(z = L/2\) is given by :

\begin{equation}\label{eq:transfer_matrix_lossy}
    \mathbf{T}_f\left(L/2\right) = \displaystyle{}\frac{\text{e}^{-\alpha_e z}}{1+C^2}\begin{pmatrix}
        1 - \gamma C^2         & C\left(1+\gamma\right) \\
        C\left(1+\gamma\right) & C^2 - \gamma
    \end{pmatrix},
\end{equation}
where \(\gamma = \text{e}^{\left(\alpha_e - \alpha_o\right)L/2}\). The forward and backward output of the optical isolator are given by: \(\mathbf{E}_{f,\;out} = \displaystyle{\frac{\left[1 - \gamma C^2 + C\left(1+\gamma\right)\right]\text{e}^{-\frac{\alpha_e L}{2}}}{\sqrt{2}\left(1+C^2\right)}}\) and \(\mathbf{E}_{b,\;out} = \displaystyle{\frac{\left[C\left(1+\gamma\right) + C^2 - \gamma \right]\text{e}^{-\frac{\alpha_e L}{2}}}{\sqrt{2}\left(1+C^2\right)}}\), respectively.\\

\begin{figure}[htbp]
    \centering
    \includegraphics[width=0.8\linewidth]{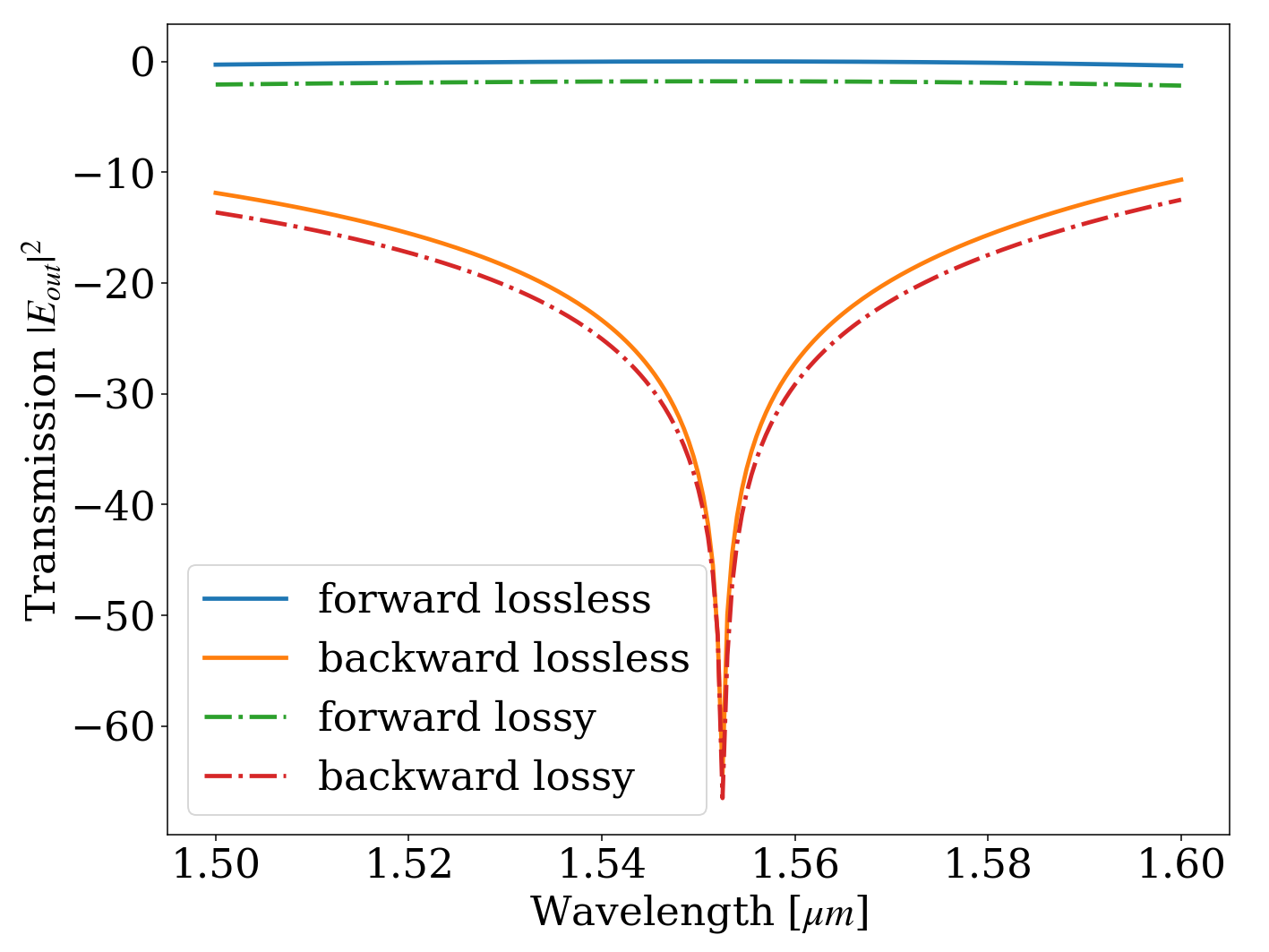}
    \caption{Simulated forward and backward transmission comparison between lossless and lossy cases.}
    \label{fig:transmission_lossy}
\end{figure}

Figure \ref{fig:transmission_lossy} depicts the forward and backward transmission in a case where the diagonal element of the permittivity of the MO material is a complex number, \(\epsilon_0 = \complexnum{4.84+i0.002}\). The loss of the supermodes caused by this imaginary part of the permittivity is approximately \qty{36}{dB\per cm}. Based on the eigenmode calculation, we can determine that \(\gamma\) is approximately \num{1.0034}. The forward and backward output intensity are then \(\num{1.001} \text{e}^{-\alpha_eL} \)and \(\num{1.4e-6}\text{e}^{-\alpha_eL}\), respectively. The isolation ratio is about \qty{-58}{dB}. In summary, although the propagation loss of \qty{36}{dB\per cm} affects the transmission loss of the optical isolator, its impact is relatively small on the isolation ratio, and the expected insertion losses are about \qty{1.8}{dB}.

\section{Conclusion}
\label{sec_conclusion}

This paper demonstrates the potential of the TMOKE coupled-waveguide system for nonreciprocal transmission. It provides a general formalism to describe the propagation within this system. In addition to the asynchronous behavior of the TMOKE coupled-waveguide system, it can be tuned to have a nonreciprocal \(\pi\) phaseshift by properly choosing its opto-geometrical parameters. This \(\pi\)-phaseshift, when combined with a \(1\times2\)  MMI, can be used to design an optical isolator. This proposed isolator shows promising performance in TE polarisation in terms of isolation ratio, insertion losses, and bandwidth. Without taking the material dispersion into account, the \qty{20}{dB}-isolation bandwidth of the isolator is numerically evaluated at \qty{35}{nm}, while its footprint measures approximately \qty{500}{\um}. Furthermore, the proposed design has three ports and can be considered a quasi-circulator, highlighting its versatile potential for miniaturizing complex optical communication, data communication, and optical sensing circuits. 


\begin{acknowledgments}
This project has received funding from the European Union’s Horizon Europe research and innovation program under grant agreement No101129645 (project CIRCULIGHT, www.circulight.eu). Views and opinions expressed are, however, those of the author(s) only and do not necessarily reflect those of the European Union or EISMEA. Neither the European Union nor the granting authority can be held responsible for them.
\end{acknowledgments}

\bibliography{references.bib}

\providecommand{\noopsort}[1]{}\providecommand{\singleletter}[1]{#1}%
\begin{thebibliography}{28}%
\makeatletter
\providecommand \@ifxundefined [1]{%
 \@ifx{#1\undefined}
}%
\providecommand \@ifnum [1]{%
 \ifnum #1\expandafter \@firstoftwo
 \else \expandafter \@secondoftwo
 \fi
}%
\providecommand \@ifx [1]{%
 \ifx #1\expandafter \@firstoftwo
 \else \expandafter \@secondoftwo
 \fi
}%
\providecommand \natexlab [1]{#1}%
\providecommand \enquote  [1]{``#1''}%
\providecommand \bibnamefont  [1]{#1}%
\providecommand \bibfnamefont [1]{#1}%
\providecommand \citenamefont [1]{#1}%
\providecommand \href@noop [0]{\@secondoftwo}%
\providecommand \href [0]{\begingroup \@sanitize@url \@href}%
\providecommand \@href[1]{\@@startlink{#1}\@@href}%
\providecommand \@@href[1]{\endgroup#1\@@endlink}%
\providecommand \@sanitize@url [0]{\catcode `\\12\catcode `\$12\catcode `\&12\catcode `\#12\catcode `\^12\catcode `\_12\catcode `\%12\relax}%
\providecommand \@@startlink[1]{}%
\providecommand \@@endlink[0]{}%
\providecommand \url  [0]{\begingroup\@sanitize@url \@url }%
\providecommand \@url [1]{\endgroup\@href {#1}{\urlprefix }}%
\providecommand \urlprefix  [0]{URL }%
\providecommand \Eprint [0]{\href }%
\providecommand \doibase [0]{https://doi.org/}%
\providecommand \selectlanguage [0]{\@gobble}%
\providecommand \bibinfo  [0]{\@secondoftwo}%
\providecommand \bibfield  [0]{\@secondoftwo}%
\providecommand \translation [1]{[#1]}%
\providecommand \BibitemOpen [0]{}%
\providecommand \bibitemStop [0]{}%
\providecommand \bibitemNoStop [0]{.\EOS\space}%
\providecommand \EOS [0]{\spacefactor3000\relax}%
\providecommand \BibitemShut  [1]{\csname bibitem#1\endcsname}%
\let\auto@bib@innerbib\@empty
\bibitem [{\citenamefont {Petermann}(1995)}]{petermanExternalOpticalFeedback1995}%
  \BibitemOpen
  \bibfield  {author} {\bibinfo {author} {\bibfnamefont {K.}~\bibnamefont {Petermann}},\ }\href@noop {} {\bibfield  {journal} {\bibinfo  {journal} {IEEE Journal of Selected Topics in Quantum Electronics}\ }\textbf {\bibinfo {volume} {1}},\ \bibinfo {pages} {480} (\bibinfo {year} {1995})}\BibitemShut {NoStop}%
\bibitem [{\citenamefont {Shekhar}\ \emph {et~al.}(2024)\citenamefont {Shekhar}, \citenamefont {Bogaerts}, \citenamefont {Chrostowski}, \citenamefont {Bowers}, \citenamefont {Hochberg}, \citenamefont {Soref},\ and\ \citenamefont {Shastri}}]{shekharRoadmappingNextGeneration2024}%
  \BibitemOpen
  \bibfield  {author} {\bibinfo {author} {\bibfnamefont {S.}~\bibnamefont {Shekhar}}, \bibinfo {author} {\bibfnamefont {W.}~\bibnamefont {Bogaerts}}, \bibinfo {author} {\bibfnamefont {L.}~\bibnamefont {Chrostowski}}, \bibinfo {author} {\bibfnamefont {J.~E.}\ \bibnamefont {Bowers}}, \bibinfo {author} {\bibfnamefont {M.}~\bibnamefont {Hochberg}}, \bibinfo {author} {\bibfnamefont {R.}~\bibnamefont {Soref}},\ and\ \bibinfo {author} {\bibfnamefont {B.~J.}\ \bibnamefont {Shastri}},\ }\href {https://doi.org/10.1038/s41467-024-44750-0} {\bibfield  {journal} {\bibinfo  {journal} {Nature Communications}\ }\textbf {\bibinfo {volume} {15}},\ \bibinfo {pages} {751} (\bibinfo {year} {2024})}\BibitemShut {NoStop}%
\bibitem [{\citenamefont {Harfouche}\ \emph {et~al.}(2020)\citenamefont {Harfouche}, \citenamefont {Kim}, \citenamefont {Wang}, \citenamefont {Santis}, \citenamefont {Zhang}, \citenamefont {Chen}, \citenamefont {Satyan}, \citenamefont {Rakuljic},\ and\ \citenamefont {Yariv}}]{harfoucheKickingHabitSemiconductor2020}%
  \BibitemOpen
  \bibfield  {author} {\bibinfo {author} {\bibfnamefont {M.}~\bibnamefont {Harfouche}}, \bibinfo {author} {\bibfnamefont {D.}~\bibnamefont {Kim}}, \bibinfo {author} {\bibfnamefont {H.}~\bibnamefont {Wang}}, \bibinfo {author} {\bibfnamefont {C.~T.}\ \bibnamefont {Santis}}, \bibinfo {author} {\bibfnamefont {Z.}~\bibnamefont {Zhang}}, \bibinfo {author} {\bibfnamefont {H.}~\bibnamefont {Chen}}, \bibinfo {author} {\bibfnamefont {N.}~\bibnamefont {Satyan}}, \bibinfo {author} {\bibfnamefont {G.}~\bibnamefont {Rakuljic}},\ and\ \bibinfo {author} {\bibfnamefont {A.}~\bibnamefont {Yariv}},\ }\href {https://doi.org/10.1364/OE.411816} {\bibfield  {journal} {\bibinfo  {journal} {Optics Express}\ }\textbf {\bibinfo {volume} {28}},\ \bibinfo {pages} {36466} (\bibinfo {year} {2020})}\BibitemShut {NoStop}%
\bibitem [{\citenamefont {Gomez}\ \emph {et~al.}(2020)\citenamefont {Gomez}, \citenamefont {Huang}, \citenamefont {Duan}, \citenamefont {Combri{\'e}}, \citenamefont {Shen}, \citenamefont {Baili}, \citenamefont {De~Rossi},\ and\ \citenamefont {Grillot}}]{gomezHighCoherenceCollapse2020}%
  \BibitemOpen
  \bibfield  {author} {\bibinfo {author} {\bibfnamefont {S.}~\bibnamefont {Gomez}}, \bibinfo {author} {\bibfnamefont {H.}~\bibnamefont {Huang}}, \bibinfo {author} {\bibfnamefont {J.}~\bibnamefont {Duan}}, \bibinfo {author} {\bibfnamefont {S.}~\bibnamefont {Combri{\'e}}}, \bibinfo {author} {\bibfnamefont {A.}~\bibnamefont {Shen}}, \bibinfo {author} {\bibfnamefont {G.}~\bibnamefont {Baili}}, \bibinfo {author} {\bibfnamefont {A.}~\bibnamefont {De~Rossi}},\ and\ \bibinfo {author} {\bibfnamefont {F.}~\bibnamefont {Grillot}},\ }\href {https://doi.org/10.1088/2515-7647/ab6a74} {\bibfield  {journal} {\bibinfo  {journal} {Journal of Physics: Photonics}\ }\textbf {\bibinfo {volume} {2}},\ \bibinfo {pages} {025005} (\bibinfo {year} {2020})}\BibitemShut {NoStop}%
\bibitem [{\citenamefont {Duan}\ \emph {et~al.}(2019)\citenamefont {Duan}, \citenamefont {Huang}, \citenamefont {Dong}, \citenamefont {Jung}, \citenamefont {Norman}, \citenamefont {Bowers},\ and\ \citenamefont {Grillot}}]{duan13$mu$Reflection2019}%
  \BibitemOpen
  \bibfield  {author} {\bibinfo {author} {\bibfnamefont {J.}~\bibnamefont {Duan}}, \bibinfo {author} {\bibfnamefont {H.}~\bibnamefont {Huang}}, \bibinfo {author} {\bibfnamefont {B.}~\bibnamefont {Dong}}, \bibinfo {author} {\bibfnamefont {D.}~\bibnamefont {Jung}}, \bibinfo {author} {\bibfnamefont {J.~C.}\ \bibnamefont {Norman}}, \bibinfo {author} {\bibfnamefont {J.~E.}\ \bibnamefont {Bowers}},\ and\ \bibinfo {author} {\bibfnamefont {F.}~\bibnamefont {Grillot}},\ }\href {https://doi.org/10.1109/LPT.2019.2895049} {\bibfield  {journal} {\bibinfo  {journal} {IEEE Photonics Technology Letters}\ }\textbf {\bibinfo {volume} {31}},\ \bibinfo {pages} {345} (\bibinfo {year} {2019})}\BibitemShut {NoStop}%
\bibitem [{\citenamefont {Huang}\ \emph {et~al.}(2024)\citenamefont {Huang}, \citenamefont {Yerkes}, \citenamefont {Su}, \citenamefont {Mehta}, \citenamefont {Cramer}, \citenamefont {O'Brien}, \citenamefont {Dehghannasiri}, \citenamefont {Dobek}, \citenamefont {Mackos}, \citenamefont {Ward}, \citenamefont {Patel}, \citenamefont {Kumar}, \citenamefont {Liu}, \citenamefont {Wu}, \citenamefont {Wang}, \citenamefont {Gao}, \citenamefont {Isenberger}, \citenamefont {Frish},\ and\ \citenamefont {Rong}}]{huangFeedbackTolerantQuantum2024}%
  \BibitemOpen
  \bibfield  {author} {\bibinfo {author} {\bibfnamefont {D.}~\bibnamefont {Huang}}, \bibinfo {author} {\bibfnamefont {S.}~\bibnamefont {Yerkes}}, \bibinfo {author} {\bibfnamefont {G.-L.}\ \bibnamefont {Su}}, \bibinfo {author} {\bibfnamefont {K.}~\bibnamefont {Mehta}}, \bibinfo {author} {\bibfnamefont {M.}~\bibnamefont {Cramer}}, \bibinfo {author} {\bibfnamefont {W.}~\bibnamefont {O'Brien}}, \bibinfo {author} {\bibfnamefont {R.}~\bibnamefont {Dehghannasiri}}, \bibinfo {author} {\bibfnamefont {S.}~\bibnamefont {Dobek}}, \bibinfo {author} {\bibfnamefont {C.}~\bibnamefont {Mackos}}, \bibinfo {author} {\bibfnamefont {T.}~\bibnamefont {Ward}}, \bibinfo {author} {\bibfnamefont {P.}~\bibnamefont {Patel}}, \bibinfo {author} {\bibfnamefont {R.}~\bibnamefont {Kumar}}, \bibinfo {author} {\bibfnamefont {S.}~\bibnamefont {Liu}}, \bibinfo {author} {\bibfnamefont {X.}~\bibnamefont {Wu}}, \bibinfo {author} {\bibfnamefont {X.}~\bibnamefont {Wang}}, \bibinfo {author} {\bibfnamefont {J.}~\bibnamefont {Gao}}, \bibinfo {author}
  {\bibfnamefont {M.}~\bibnamefont {Isenberger}}, \bibinfo {author} {\bibfnamefont {H.}~\bibnamefont {Frish}},\ and\ \bibinfo {author} {\bibfnamefont {H.}~\bibnamefont {Rong}},\ }in\ \href {https://doi.org/10.1364/OFC.2024.M3C.2} {\emph {\bibinfo {booktitle} {Optical {{Fiber Communication Conference}} ({{OFC}}) 2024}}}\ (\bibinfo  {publisher} {Optica Publishing Group},\ \bibinfo {address} {San Diego California},\ \bibinfo {year} {2024})\ p.\ \bibinfo {pages} {M3C.2}\BibitemShut {NoStop}%
\bibitem [{\citenamefont {Tian}\ \emph {et~al.}(2021)\citenamefont {Tian}, \citenamefont {Liu}, \citenamefont {Siddharth}, \citenamefont {Wang}, \citenamefont {Bl{\'e}sin}, \citenamefont {He}, \citenamefont {Kippenberg},\ and\ \citenamefont {Bhave}}]{tianMagneticFreeSiliconNitride2021}%
  \BibitemOpen
  \bibfield  {author} {\bibinfo {author} {\bibfnamefont {H.}~\bibnamefont {Tian}}, \bibinfo {author} {\bibfnamefont {J.}~\bibnamefont {Liu}}, \bibinfo {author} {\bibfnamefont {A.}~\bibnamefont {Siddharth}}, \bibinfo {author} {\bibfnamefont {R.~N.}\ \bibnamefont {Wang}}, \bibinfo {author} {\bibfnamefont {T.}~\bibnamefont {Bl{\'e}sin}}, \bibinfo {author} {\bibfnamefont {J.}~\bibnamefont {He}}, \bibinfo {author} {\bibfnamefont {T.~J.}\ \bibnamefont {Kippenberg}},\ and\ \bibinfo {author} {\bibfnamefont {S.~A.}\ \bibnamefont {Bhave}},\ }\href@noop {} {\bibfield  {journal} {\bibinfo  {journal} {arXiv:2104.01158 [physics]}\ } (\bibinfo {year} {2021})},\ \Eprint {https://arxiv.org/abs/2104.01158} {arXiv:2104.01158 [physics]} \BibitemShut {NoStop}%
\bibitem [{\citenamefont {Shoman}\ \emph {et~al.}(2021)\citenamefont {Shoman}, \citenamefont {Jaeger}, \citenamefont {Mosquera}, \citenamefont {Jayatilleka}, \citenamefont {Ma}, \citenamefont {Rong}, \citenamefont {Shekhar},\ and\ \citenamefont {Chrostowski}}]{shomanStableReducedLinewidthLaser2021a}%
  \BibitemOpen
  \bibfield  {author} {\bibinfo {author} {\bibfnamefont {H.}~\bibnamefont {Shoman}}, \bibinfo {author} {\bibfnamefont {N.~A.~F.}\ \bibnamefont {Jaeger}}, \bibinfo {author} {\bibfnamefont {C.}~\bibnamefont {Mosquera}}, \bibinfo {author} {\bibfnamefont {H.}~\bibnamefont {Jayatilleka}}, \bibinfo {author} {\bibfnamefont {M.}~\bibnamefont {Ma}}, \bibinfo {author} {\bibfnamefont {H.}~\bibnamefont {Rong}}, \bibinfo {author} {\bibfnamefont {S.}~\bibnamefont {Shekhar}},\ and\ \bibinfo {author} {\bibfnamefont {L.}~\bibnamefont {Chrostowski}},\ }\href {https://doi.org/10.1109/JLT.2021.3096460} {\bibfield  {journal} {\bibinfo  {journal} {Journal of Lightwave Technology}\ }\textbf {\bibinfo {volume} {39}},\ \bibinfo {pages} {6215} (\bibinfo {year} {2021})}\BibitemShut {NoStop}%
\bibitem [{\citenamefont {Hutchings}\ \emph {et~al.}(2013)\citenamefont {Hutchings}, \citenamefont {Holmes}, \citenamefont {Zhang}, \citenamefont {Dulal}, \citenamefont {Block}, \citenamefont {Sung}, \citenamefont {Seaton},\ and\ \citenamefont {Stadler}}]{hutchingsQuasiPhaseMatchedFaradayRotation2013}%
  \BibitemOpen
  \bibfield  {author} {\bibinfo {author} {\bibfnamefont {D.~C.}\ \bibnamefont {Hutchings}}, \bibinfo {author} {\bibfnamefont {B.~M.}\ \bibnamefont {Holmes}}, \bibinfo {author} {\bibfnamefont {C.}~\bibnamefont {Zhang}}, \bibinfo {author} {\bibfnamefont {P.}~\bibnamefont {Dulal}}, \bibinfo {author} {\bibfnamefont {A.~D.}\ \bibnamefont {Block}}, \bibinfo {author} {\bibfnamefont {S.-Y.}\ \bibnamefont {Sung}}, \bibinfo {author} {\bibfnamefont {N.~C.~A.}\ \bibnamefont {Seaton}},\ and\ \bibinfo {author} {\bibfnamefont {B.~J.~H.}\ \bibnamefont {Stadler}},\ }\href {https://doi.org/10.1109/JPHOT.2013.2292339} {\bibfield  {journal} {\bibinfo  {journal} {IEEE Photonics Journal}\ }\textbf {\bibinfo {volume} {5}},\ \bibinfo {pages} {6602512} (\bibinfo {year} {2013})}\BibitemShut {NoStop}%
\bibitem [{\citenamefont {Pintus}\ \emph {et~al.}(2011)\citenamefont {Pintus}, \citenamefont {Tien},\ and\ \citenamefont {Bowers}}]{pintusDesignMagnetoOpticalRing2011}%
  \BibitemOpen
  \bibfield  {author} {\bibinfo {author} {\bibfnamefont {P.}~\bibnamefont {Pintus}}, \bibinfo {author} {\bibfnamefont {M.-C.}\ \bibnamefont {Tien}},\ and\ \bibinfo {author} {\bibfnamefont {J.~E.}\ \bibnamefont {Bowers}},\ }\href {https://doi.org/10.1109/LPT.2011.2164397} {\bibfield  {journal} {\bibinfo  {journal} {IEEE Photonics Technology Letters}\ }\textbf {\bibinfo {volume} {23}},\ \bibinfo {pages} {1670} (\bibinfo {year} {2011})}\BibitemShut {NoStop}%
\bibitem [{\citenamefont {Huang}\ \emph {et~al.}(2016)\citenamefont {Huang}, \citenamefont {Pintus}, \citenamefont {Zhang}, \citenamefont {Shoji}, \citenamefont {Mizumoto},\ and\ \citenamefont {Bowers}}]{huangElectricallyDrivenThermally2016}%
  \BibitemOpen
  \bibfield  {author} {\bibinfo {author} {\bibfnamefont {D.}~\bibnamefont {Huang}}, \bibinfo {author} {\bibfnamefont {P.}~\bibnamefont {Pintus}}, \bibinfo {author} {\bibfnamefont {C.}~\bibnamefont {Zhang}}, \bibinfo {author} {\bibfnamefont {Y.}~\bibnamefont {Shoji}}, \bibinfo {author} {\bibfnamefont {T.}~\bibnamefont {Mizumoto}},\ and\ \bibinfo {author} {\bibfnamefont {J.~E.}\ \bibnamefont {Bowers}},\ }\href {https://doi.org/10.1109/JSTQE.2016.2588778} {\bibfield  {journal} {\bibinfo  {journal} {IEEE Journal of Selected Topics in Quantum Electronics}\ }\textbf {\bibinfo {volume} {22}},\ \bibinfo {pages} {271} (\bibinfo {year} {2016})}\BibitemShut {NoStop}%
\bibitem [{\citenamefont {Shoji}\ \emph {et~al.}(2014)\citenamefont {Shoji}, \citenamefont {Shirato},\ and\ \citenamefont {Mizumoto}}]{shojiSiliconMachZehnder2014}%
  \BibitemOpen
  \bibfield  {author} {\bibinfo {author} {\bibfnamefont {Y.}~\bibnamefont {Shoji}}, \bibinfo {author} {\bibfnamefont {Y.}~\bibnamefont {Shirato}},\ and\ \bibinfo {author} {\bibfnamefont {T.}~\bibnamefont {Mizumoto}},\ }\href {https://doi.org/10.7567/JJAP.53.022202} {\bibfield  {journal} {\bibinfo  {journal} {Japanese Journal of Applied Physics}\ }\textbf {\bibinfo {volume} {53}},\ \bibinfo {pages} {022202} (\bibinfo {year} {2014})}\BibitemShut {NoStop}%
\bibitem [{\citenamefont {Huang}\ \emph {et~al.}(2017)\citenamefont {Huang}, \citenamefont {Pintus}, \citenamefont {Shoji}, \citenamefont {Morton}, \citenamefont {Mizumoto},\ and\ \citenamefont {Bowers}}]{huangIntegratedBroadbandCe2017}%
  \BibitemOpen
  \bibfield  {author} {\bibinfo {author} {\bibfnamefont {D.}~\bibnamefont {Huang}}, \bibinfo {author} {\bibfnamefont {P.}~\bibnamefont {Pintus}}, \bibinfo {author} {\bibfnamefont {Y.}~\bibnamefont {Shoji}}, \bibinfo {author} {\bibfnamefont {P.}~\bibnamefont {Morton}}, \bibinfo {author} {\bibfnamefont {T.}~\bibnamefont {Mizumoto}},\ and\ \bibinfo {author} {\bibfnamefont {J.~E.}\ \bibnamefont {Bowers}},\ }\href {https://doi.org/10.1364/OL.42.004901} {\bibfield  {journal} {\bibinfo  {journal} {Optics Letters}\ }\textbf {\bibinfo {volume} {42}},\ \bibinfo {pages} {4901} (\bibinfo {year} {2017})}\BibitemShut {NoStop}%
\bibitem [{\citenamefont {Zhang}\ \emph {et~al.}(2017)\citenamefont {Zhang}, \citenamefont {Dulal}, \citenamefont {Stadler},\ and\ \citenamefont {Hutchings}}]{zhangMonolithicallyIntegratedTEmode1D2017}%
  \BibitemOpen
  \bibfield  {author} {\bibinfo {author} {\bibfnamefont {C.}~\bibnamefont {Zhang}}, \bibinfo {author} {\bibfnamefont {P.}~\bibnamefont {Dulal}}, \bibinfo {author} {\bibfnamefont {B.~J.~H.}\ \bibnamefont {Stadler}},\ and\ \bibinfo {author} {\bibfnamefont {D.~C.}\ \bibnamefont {Hutchings}},\ }\href {https://doi.org/10.1038/s41598-017-06043-z} {\bibfield  {journal} {\bibinfo  {journal} {Scientific Reports}\ }\textbf {\bibinfo {volume} {7}},\ \bibinfo {pages} {5820} (\bibinfo {year} {2017})}\BibitemShut {NoStop}%
\bibitem [{\citenamefont {Srinivasan}\ \emph {et~al.}(2019)\citenamefont {Srinivasan}, \citenamefont {Zhang}, \citenamefont {Dulal}, \citenamefont {Radu}, \citenamefont {Gage}, \citenamefont {Hutchings},\ and\ \citenamefont {Stadler}}]{srinivasanHighGyrotropySeedlayerFreeCe2019}%
  \BibitemOpen
  \bibfield  {author} {\bibinfo {author} {\bibfnamefont {K.}~\bibnamefont {Srinivasan}}, \bibinfo {author} {\bibfnamefont {C.}~\bibnamefont {Zhang}}, \bibinfo {author} {\bibfnamefont {P.}~\bibnamefont {Dulal}}, \bibinfo {author} {\bibfnamefont {C.}~\bibnamefont {Radu}}, \bibinfo {author} {\bibfnamefont {T.~E.}\ \bibnamefont {Gage}}, \bibinfo {author} {\bibfnamefont {D.~C.}\ \bibnamefont {Hutchings}},\ and\ \bibinfo {author} {\bibfnamefont {B.~J.~H.}\ \bibnamefont {Stadler}},\ }\href {https://doi.org/10.1021/acsphotonics.9b00707} {\bibfield  {journal} {\bibinfo  {journal} {ACS Photonics}\ }\textbf {\bibinfo {volume} {6}},\ \bibinfo {pages} {2455} (\bibinfo {year} {2019})}\BibitemShut {NoStop}%
\bibitem [{\citenamefont {Liu}\ \emph {et~al.}(2022)\citenamefont {Liu}, \citenamefont {Shoji},\ and\ \citenamefont {Mizumoto}}]{liuTEmodeMagnetoopticalIsolator2022}%
  \BibitemOpen
  \bibfield  {author} {\bibinfo {author} {\bibfnamefont {S.}~\bibnamefont {Liu}}, \bibinfo {author} {\bibfnamefont {Y.}~\bibnamefont {Shoji}},\ and\ \bibinfo {author} {\bibfnamefont {T.}~\bibnamefont {Mizumoto}},\ }\href {https://doi.org/10.1364/OE.454751} {\bibfield  {journal} {\bibinfo  {journal} {Optics Express}\ }\textbf {\bibinfo {volume} {30}},\ \bibinfo {pages} {9934} (\bibinfo {year} {2022})}\BibitemShut {NoStop}%
\bibitem [{\citenamefont {Yan}\ \emph {et~al.}(2020)\citenamefont {Yan}, \citenamefont {Yang}, \citenamefont {Liu}, \citenamefont {Zhang}, \citenamefont {Xia}, \citenamefont {Kang}, \citenamefont {Yang}, \citenamefont {Qin}, \citenamefont {Deng},\ and\ \citenamefont {Bi}}]{yanWaveguideintegratedHighperformanceMagnetooptical2020}%
  \BibitemOpen
  \bibfield  {author} {\bibinfo {author} {\bibfnamefont {W.}~\bibnamefont {Yan}}, \bibinfo {author} {\bibfnamefont {Y.}~\bibnamefont {Yang}}, \bibinfo {author} {\bibfnamefont {S.}~\bibnamefont {Liu}}, \bibinfo {author} {\bibfnamefont {Y.}~\bibnamefont {Zhang}}, \bibinfo {author} {\bibfnamefont {S.}~\bibnamefont {Xia}}, \bibinfo {author} {\bibfnamefont {T.}~\bibnamefont {Kang}}, \bibinfo {author} {\bibfnamefont {W.}~\bibnamefont {Yang}}, \bibinfo {author} {\bibfnamefont {J.}~\bibnamefont {Qin}}, \bibinfo {author} {\bibfnamefont {L.}~\bibnamefont {Deng}},\ and\ \bibinfo {author} {\bibfnamefont {L.}~\bibnamefont {Bi}},\ }\href {https://doi.org/10.1364/OPTICA.408458} {\bibfield  {journal} {\bibinfo  {journal} {Optica}\ }\textbf {\bibinfo {volume} {7}},\ \bibinfo {pages} {1555} (\bibinfo {year} {2020})}\BibitemShut {NoStop}%
\bibitem [{\citenamefont {Yamaguchi}\ \emph {et~al.}(2018)\citenamefont {Yamaguchi}, \citenamefont {Shoji},\ and\ \citenamefont {Mizumoto}}]{yamaguchiLowlossWaveguideOptical2018}%
  \BibitemOpen
  \bibfield  {author} {\bibinfo {author} {\bibfnamefont {R.}~\bibnamefont {Yamaguchi}}, \bibinfo {author} {\bibfnamefont {Y.}~\bibnamefont {Shoji}},\ and\ \bibinfo {author} {\bibfnamefont {T.}~\bibnamefont {Mizumoto}},\ }\href {https://doi.org/10.1364/OE.26.021271} {\bibfield  {journal} {\bibinfo  {journal} {Optics Express}\ }\textbf {\bibinfo {volume} {26}},\ \bibinfo {pages} {21271} (\bibinfo {year} {2018})}\BibitemShut {NoStop}%
\bibitem [{\citenamefont {Zhang}\ \emph {et~al.}(2019)\citenamefont {Zhang}, \citenamefont {Du}, \citenamefont {Wang}, \citenamefont {Fakhrul}, \citenamefont {Liu}, \citenamefont {Deng}, \citenamefont {Huang}, \citenamefont {Pintus}, \citenamefont {Bowers}, \citenamefont {Ross}, \citenamefont {Hu},\ and\ \citenamefont {Bi}}]{zhangMonolithicIntegrationBroadband2019}%
  \BibitemOpen
  \bibfield  {author} {\bibinfo {author} {\bibfnamefont {Y.}~\bibnamefont {Zhang}}, \bibinfo {author} {\bibfnamefont {Q.}~\bibnamefont {Du}}, \bibinfo {author} {\bibfnamefont {C.}~\bibnamefont {Wang}}, \bibinfo {author} {\bibfnamefont {T.}~\bibnamefont {Fakhrul}}, \bibinfo {author} {\bibfnamefont {S.}~\bibnamefont {Liu}}, \bibinfo {author} {\bibfnamefont {L.}~\bibnamefont {Deng}}, \bibinfo {author} {\bibfnamefont {D.}~\bibnamefont {Huang}}, \bibinfo {author} {\bibfnamefont {P.}~\bibnamefont {Pintus}}, \bibinfo {author} {\bibfnamefont {J.}~\bibnamefont {Bowers}}, \bibinfo {author} {\bibfnamefont {C.~A.}\ \bibnamefont {Ross}}, \bibinfo {author} {\bibfnamefont {J.}~\bibnamefont {Hu}},\ and\ \bibinfo {author} {\bibfnamefont {L.}~\bibnamefont {Bi}},\ }\href {https://doi.org/10.1364/OPTICA.6.000473} {\bibfield  {journal} {\bibinfo  {journal} {Optica}\ }\textbf {\bibinfo {volume} {6}},\ \bibinfo {pages} {473} (\bibinfo {year} {2019})}\BibitemShut {NoStop}%
\bibitem [{\citenamefont {Ma}\ \emph {et~al.}(2021)\citenamefont {Ma}, \citenamefont {Reniers}, \citenamefont {Shoji}, \citenamefont {Mizumoto}, \citenamefont {Williams}, \citenamefont {Jiao},\ and\ \citenamefont {{van der Tol}}}]{maIntegratedPolarizationindependentOptical2021}%
  \BibitemOpen
  \bibfield  {author} {\bibinfo {author} {\bibfnamefont {R.}~\bibnamefont {Ma}}, \bibinfo {author} {\bibfnamefont {S.}~\bibnamefont {Reniers}}, \bibinfo {author} {\bibfnamefont {Y.}~\bibnamefont {Shoji}}, \bibinfo {author} {\bibfnamefont {T.}~\bibnamefont {Mizumoto}}, \bibinfo {author} {\bibfnamefont {K.}~\bibnamefont {Williams}}, \bibinfo {author} {\bibfnamefont {Y.}~\bibnamefont {Jiao}},\ and\ \bibinfo {author} {\bibfnamefont {J.}~\bibnamefont {{van der Tol}}},\ }\href {https://doi.org/10.1364/OPTICA.443097} {\bibfield  {journal} {\bibinfo  {journal} {Optica}\ }\textbf {\bibinfo {volume} {8}},\ \bibinfo {pages} {1654} (\bibinfo {year} {2021})}\BibitemShut {NoStop}%
\bibitem [{\citenamefont {Abadian}\ \emph {et~al.}(2021)\citenamefont {Abadian}, \citenamefont {Magno}, \citenamefont {Yam},\ and\ \citenamefont {Dagens}}]{abadianBroadbandPlasmonicIsolator2021}%
  \BibitemOpen
  \bibfield  {author} {\bibinfo {author} {\bibfnamefont {S.}~\bibnamefont {Abadian}}, \bibinfo {author} {\bibfnamefont {G.}~\bibnamefont {Magno}}, \bibinfo {author} {\bibfnamefont {V.}~\bibnamefont {Yam}},\ and\ \bibinfo {author} {\bibfnamefont {B.}~\bibnamefont {Dagens}},\ }\href {https://doi.org/10.1364/OE.415969} {\bibfield  {journal} {\bibinfo  {journal} {Optics Express}\ }\textbf {\bibinfo {volume} {29}},\ \bibinfo {pages} {4091} (\bibinfo {year} {2021})}\BibitemShut {NoStop}%
\bibitem [{\citenamefont {Ho}\ \emph {et~al.}(2018)\citenamefont {Ho}, \citenamefont {Im}, \citenamefont {Pae}, \citenamefont {Ri}, \citenamefont {Han},\ and\ \citenamefont {Herrmann}}]{hoSwitchablePlasmonicRouters2018}%
  \BibitemOpen
  \bibfield  {author} {\bibinfo {author} {\bibfnamefont {K.-S.}\ \bibnamefont {Ho}}, \bibinfo {author} {\bibfnamefont {S.-J.}\ \bibnamefont {Im}}, \bibinfo {author} {\bibfnamefont {J.-S.}\ \bibnamefont {Pae}}, \bibinfo {author} {\bibfnamefont {C.-S.}\ \bibnamefont {Ri}}, \bibinfo {author} {\bibfnamefont {Y.-H.}\ \bibnamefont {Han}},\ and\ \bibinfo {author} {\bibfnamefont {J.}~\bibnamefont {Herrmann}},\ }\href {https://doi.org/10.1038/s41598-018-28567-8} {\bibfield  {journal} {\bibinfo  {journal} {Scientific Reports}\ }\textbf {\bibinfo {volume} {8}},\ \bibinfo {pages} {10584} (\bibinfo {year} {2018})}\BibitemShut {NoStop}%
\bibitem [{\citenamefont {Zhuromskyy}\ \emph {et~al.}(1999)\citenamefont {Zhuromskyy}, \citenamefont {Lohmeyer}, \citenamefont {Bahlmann}, \citenamefont {Dotsch}, \citenamefont {Hertel},\ and\ \citenamefont {Popkov}}]{zhuromskyyAnalysisPolarizationIndependent1999}%
  \BibitemOpen
  \bibfield  {author} {\bibinfo {author} {\bibfnamefont {O.}~\bibnamefont {Zhuromskyy}}, \bibinfo {author} {\bibfnamefont {M.}~\bibnamefont {Lohmeyer}}, \bibinfo {author} {\bibfnamefont {N.}~\bibnamefont {Bahlmann}}, \bibinfo {author} {\bibfnamefont {H.}~\bibnamefont {Dotsch}}, \bibinfo {author} {\bibfnamefont {P.}~\bibnamefont {Hertel}},\ and\ \bibinfo {author} {\bibfnamefont {A.}~\bibnamefont {Popkov}},\ }\href {https://doi.org/10.1109/50.774255} {\bibfield  {journal} {\bibinfo  {journal} {Journal of Lightwave Technology}\ }\textbf {\bibinfo {volume} {17}},\ \bibinfo {pages} {1200} (\bibinfo {year} {1999})}\BibitemShut {NoStop}%
\bibitem [{\citenamefont {Chuang}(2012)}]{chuang2012physics}%
  \BibitemOpen
  \bibfield  {author} {\bibinfo {author} {\bibfnamefont {S.~L.}\ \bibnamefont {Chuang}},\ }\href@noop {} {\emph {\bibinfo {title} {Physics of photonic devices}}}\ (\bibinfo  {publisher} {John Wiley \& Sons},\ \bibinfo {year} {2012})\BibitemShut {NoStop}%
\bibitem [{\citenamefont {Yariv}(1973)}]{yariv1973coupled}%
  \BibitemOpen
  \bibfield  {author} {\bibinfo {author} {\bibfnamefont {A.}~\bibnamefont {Yariv}},\ }\href@noop {} {\bibfield  {journal} {\bibinfo  {journal} {IEEE Journal of Quantum Electronics}\ }\textbf {\bibinfo {volume} {9}},\ \bibinfo {pages} {919} (\bibinfo {year} {1973})}\BibitemShut {NoStop}%
\bibitem [{\citenamefont {Yamamoto}\ and\ \citenamefont {Makimoto}(1974)}]{yamamoto1974circuit}%
  \BibitemOpen
  \bibfield  {author} {\bibinfo {author} {\bibfnamefont {S.}~\bibnamefont {Yamamoto}}\ and\ \bibinfo {author} {\bibfnamefont {T.}~\bibnamefont {Makimoto}},\ }\href@noop {} {\bibfield  {journal} {\bibinfo  {journal} {Journal of Applied Physics}\ }\textbf {\bibinfo {volume} {45}},\ \bibinfo {pages} {882} (\bibinfo {year} {1974})}\BibitemShut {NoStop}%
\bibitem [{\citenamefont {Soldano}\ and\ \citenamefont {Pennings}(1995)}]{soldanoOpticalMultimodeInterference1995}%
  \BibitemOpen
  \bibfield  {author} {\bibinfo {author} {\bibfnamefont {L.}~\bibnamefont {Soldano}}\ and\ \bibinfo {author} {\bibfnamefont {E.}~\bibnamefont {Pennings}},\ }\href {https://doi.org/10.1109/50.372474} {\bibfield  {journal} {\bibinfo  {journal} {Journal of Lightwave Technology}\ }\textbf {\bibinfo {volume} {13}},\ \bibinfo {pages} {615} (\bibinfo {year} {1995})}\BibitemShut {NoStop}%
\bibitem [{\citenamefont {Srinivasan}\ and\ \citenamefont {Stadler}(2022)}]{srinivasanReviewIntegratedMagnetooptical2022}%
  \BibitemOpen
  \bibfield  {author} {\bibinfo {author} {\bibfnamefont {K.}~\bibnamefont {Srinivasan}}\ and\ \bibinfo {author} {\bibfnamefont {B.~J.~H.}\ \bibnamefont {Stadler}},\ }\href {https://doi.org/10.1364/OME.447398} {\bibfield  {journal} {\bibinfo  {journal} {Optical Materials Express}\ }\textbf {\bibinfo {volume} {12}},\ \bibinfo {pages} {697} (\bibinfo {year} {2022})}\BibitemShut {NoStop}%
\end{thebibliography}%
\end{document}